\documentclass[]{pasj01}
\usepackage[switch,mathlines]{lineno}

\Received{}
\Accepted{}
 
\begin{document} 

\title{ 
Evolution of grain size distribution in the
circum-galactic medium}

\author{Hiroyuki \textsc{Hirashita}\altaffilmark{1,2}
}
\altaffiltext{1}{Institute of Astronomy and Astrophysics, Academia Sinica,
Astronomy-Mathematics Building, No.\ 1, Section 4,
Roosevelt Road, Taipei 10617, Taiwan}
\altaffiltext{2}{Theoretical Astrophysics, Department of Earth and Space Science, Osaka University,
1-1 Machikaneyama, Toyonaka, Osaka 560-0043, Japan}
\email{hirashita@asiaa.sinica.edu.tw}




\maketitle

\begin{abstract}
In order to theoretically understand dust properties in the circum-galactic medium (CGM),
we construct a dust evolution model that incorporates the evolution of grain size
distribution. We treat each of the galaxy and the CGM as a one-zone
object, and consider the mass exchange between them.
We take into account dust production and interstellar dust processing for the galaxy based on our
previous models, and newly incorporate sputtering in the hot phase and shattering in
the cool phase for the CGM.
We find that shattering increases the dust destruction (sputtering) efficiency in the CGM.
The functional shape of the grain size distribution in the CGM evolves following
that in the galaxy, but it is sensitive to
the balance between sputtering and shattering in the CGM.
For an observational test, we discuss the wavelength dependence of the reddening in the CGM
traced by background quasar colors, arguing that, in order to explain the observed reddening
level, a rapid inflow from the CGM
to the galaxy is favored because of quick dust/metal enrichment.
Small grain production by shattering in the CGM also helps
to explain the rise of dust extinction toward short wavelengths.
\end{abstract}

\section{Introduction}

Dust affects the brightness of background objects through extinction, that is,
scattering and absorption of light. Dust usually causes reddening, since
it absorbs and scatters light more efficiently at shorter wavelengths
(e.g., \cite{Draine:2003ac}).
Dust is prevailing in the interstellar medium (ISM); thus, in order to
estimate the intrinsic luminosity or color of a Galactic object,
we need to correct the observed flux for the dust extinction in the line
of sight.
Observation of an extragalactic object is also affected by the reddening in the ISM of
its host galaxy.
Thus, it is important to understand the dust extinction properties in the ISM of
the Galaxy and extragalactic objects.

Dust is also known to be distributed on larger scales than galaxies, that is,
in the intergalactic medium (IGM) or in the circum-galalctic medium (CGM).
Background quasars are used to trace the extinction
of foreground absorption line systems distributed in the circum-galactic or intergalactic space.
Since dust extinction in these systems is generally small, reddening of
the IGM or CGM has been detected in a statistical manner through a large sample of quasars
observed mainly by the Sloan Digital Sky Survey (SDSS; \cite{York:2000aa}).
For example, \citet{York:2006aa} obtained the dust extinction curves of foreground
absorption systems from the comparison between quasar samples
with and without intervening absorbers.
\citet{Menard:2010aa}
detected reddening on a scale of several Mpc around galaxies
at median $z\sim 0.3$ ($z$ is the redshift) by taking a
cross-correlation between the color excess of background quasars and the projected density
of galaxies.
A similar radial reddening profile was obtained
for nearer ($z\sim 0.05$) galaxies by
\citet{Peek:2015aa}.
\citet{Menard:2012aa} detected reddening in Mg \textsc{ii} absorbers, which are
considered to be located in the
CGM in the intervening systems (e.g., \cite{Steidel:1994aa,Lan:2020aa}).
In addition to the above absorption studies, \citet{Meinke:2023aa} found
extended dust emissions around a sample of galaxies at $z\sim 1$
by stacking millimeter data.
An analytic cosmological galaxy structure model developed by \citet{Masaki:2012aa} also supported
such a large-scale dust distribution.

Dust in the CGM and IGM is important in the following aspects.
Dust in the CGM contributes significantly to the total dust content in the Universe,
since galaxy halos contain an amount of dust comparable to galaxy disks on average
\citep{Menard:2010aa,Fukugita:2011aa}.
As mentioned above, dust distributed in a large volume of the Universe causes slight
extinction or reddening of background sources, causing some bias in the colors
and fluxes of the objects used as statistical tracers of large-scale structures
\citep{Mortsell:2003aa,Avgoustidis:2009aa}.
In addition to reddening, dust could also contribute to the cooling of the hot
($T_\mathrm{gas}\gtrsim 10^6$ K, where $T_\mathrm{gas}$ is the gas temperature) gas
contained in the CGM (e.g., \cite{Dwek:1987aa,Tsai:1995aa,Mathews:2003aa}).
In cooler gas in the CGM and IGM, dust could play an important role in gas heating
through the photoelectric effect
\citep{Inoue:2003aa,Inoue:2004aa}.
Therefore, in order  to reveal the thermal properties of the CGM and IGM,
the evolution of dust needs to be clarified.

The dust in the CGM should be supplied from the central galaxy.
There are some observations that showed spatial dust distributions extending
the galactic disk to the CGM, indicating a connection between the dust in these
two regions \citep{Howk:1997aa,Alton:1998aa}.
Galactic outflows driven by energy input (feedback) from
supernovae (SNe) and active galactic nuclei
(AGNs; e.g., \cite{Veilleux:2005aa}) can transport dust from the galaxy to the CGM.
Some hydrodynamic simulations showed that this dust transport actually
happens \citep{Zu:2011aa,McKinnon:2016aa,Hou:2017aa,Aoyama:2018aa}.
{A detailed high-resolution simulation showed that a significant fraction of the dust
survives against sputtering in galactic outflows \citep{Richie:2024aa}.
Semi-analytic frameworks also provide theoretically expected dust mass budget in the
CGM \citep{Triani:2020aa,Parente:2023aa}.}
Recent theoretical studies also suggested that galactic outflows can be an important
mechanism for the loss of dust from galaxies at high redshift
\citep{Burgarella:2020aa,Nanni:2020aa}.
Radiation pressure from stars provides another way of supplying dust to the CGM \citep{Ferrara:1991aa,Davies:1998aa,Bianchi:2005aa,Hirashita:2019ab}.
The CGM is also a reservoir of gas for the central galaxy.
The infall of the CGM into the galaxy can trigger star formation, which further
enhances stellar feedback.
Therefore, the mass exchange between the CGM and its central galaxy is the key
process that clarifies the origin of the dust in the CGM.

\citet{Otsuki:2024aa} (hereafter OH24)
developed a dust enrichment model in the CGM by considering the mass exchange
with the central galaxy. The evolution of the dust content in the galaxy is calculated
in a manner consistent with the metal enrichment based on
an already developed model
(e.g., \cite{Lisenfeld:1998aa,Dwek:1998aa,Hirashita:1999aa}).
The model includes stellar dust production,
dust growth in the ISM, and dust destruction in SN shocks.
The dust content in the CGM is calculated by considering dust supply from galactic outflows
and dust destruction by sputtering.
The latter process occurs in the hot ($T_\mathrm{gas}\gtrsim 10^6$ K) phase
\citep{Tsai:1995aa} of the CGM.
OH24 focused on the dust abundance at ages comparable to the present-day Universe
($t\sim 10$ Gyr), and found that the resulting dust mass in the CGM is
consistent with the value derived from a large sample of SDSS galaxies.
{Using a much more comprehensive model based on a semi-analytic framework,
\citet{Parente:2023aa} also obtained a similar CGM dust mass, confirming that the
physical processes regarding the CGM dust evolution are successfully included in the models.
In particular,} the dust mass in the CGM can be explained by the scenario in which the dust
is supplied by the outflows from the central galaxy.

However, there is another important aspect for the dust in the CGM, which is
the grain size distribution (the distribution function of grain size). In fact,
the above important processes of dust in the IGM and CGM, that is, reddening and
heating/cooling occur in a way dependent on the grain size,
more precisely the grain size distribution. Thus, for a detailed understanding of
these processes,
it is crucial to clarify not only the total dust abundance but also the grain size
distribution. In this paper, we focus on the CGM,
since it is the site where the dust produced in the galaxy is directly injected.
Therefore, the purpose of this paper is to construct a model that describes the
evolution of the grain size distribution in the CGM.
To achieve this goal, we need to treat the evolution of grain size distribution in the central
galaxy, with which the CGM exchanges mass.
The grain size distribution in a galaxy is already modeled in
some previous studies \citep{Asano:2013aa,Hirashita:2019aa}, which we are able to
utilize to model the central galaxy in this paper.
We extend the model to include the dust evolution in the CGM.

In addition to the dust destruction by sputtering, there is a process that is potentially important
for the evolution of the grain size distribution in the CGM:
\citet{Hirashita:2021ab} showed that shattering could occur in cool clumps in the CGM.
Coexistence of cool clumps and hot gas is also shown by
a high-resolution cosmological simulation \citep{Ramesh:2023aa}.
In cool clumps, grain motion, which causes grain--grain collisions,
is assumed to be governed by turbulence.
Some observations showed that the cool medium in the CGM is turbulent
\citep{Qu:2022aa,Chen:2023aa}.
Shattering preserves the total dust mass, so that it was not explicitly included in OH24.
However, shattering is considered to be a unique process that could explain the
existence of small grains indicated by the rest-frame ultraviolet excess in the reddening curves
\citep{Hirashita:2021ab}.
The above cool phase could be traced by Mg\,\textsc{ii} absorbers, which
\citet{Lan:2017aa} considered to have a typical gas density of
$n_\mathrm{H}\sim 0.3$~cm$^{-3}$ ($n_\mathrm{H}$ is the hydrogen number density)
and a typical dimension of 30~pc.
Gas clumps of similar size are also observed in the Milky Way halo
(e.g., \cite{BenBekhti:2009aa})
as well as in quasar absorption line systems
at high redshift (e.g., \cite{Rauch:1999aa,Prochaska:2009aa}).
\citet{Lan:2019aa}, based on an
analytically estimated evaporation time-scale, found that the lifetime of a cool clump
is $\mbox{a few}\times 10^8$ yr.
\citet{Hirashita:2021ab} showed that shattering in cool clumps with a duration of
$\mbox{a few}\times 10^8$ yr raises the abundance of small grains
to a level high enough to
explain the ultraviolet excess in the reddening curves mentioned above.

In this paper, we construct a theoretical framework for the evolution of
grain size distribution in the CGM by including both shattering and sputtering.
We extend the model developed by OH24 to include the evolution of grain size distribution.
As done by OH24,
we treat the CGM in a consistent manner with its central galaxy by including
mass exchange.
{As emphasized by OH24, we take an analytical approach without solving hydrodynamical
evolution. This serves to directly address how the grain size distribution in the CGM is
affected by the efficiencies or time-scales of
relevant processes. In other words,}
the model developed in this paper will clarify what determines the grain size distribution
in the CGM. We also calculate reddening curves in order
to directly output observable properties of the CGM dust.

This paper is organized as follows.
In section~\ref{sec:model}, we describe the evolution model of grain size distribution.
In section~\ref{sec:result}, we show the results. We further use these results for
extended predictions and discussion in section \ref{sec:discussion}.
Finally we give conclusions in section \ref{sec:conclusion}.

\section{Model}\label{sec:model}

We construct a model that describes the evolution of grain size distribution in a galaxy
and its CGM.
Since the dust in the CGM is supplied from the central galaxy,
we also need to model the mass exchange between the CGM and the galaxy as well
as the dust evolution within the galaxy.
OH24 already modeled the evolution of dust mass in the CGM by considering
the mass exchange with the galaxy and including relevant processes
that affect the dust content in the galaxy and the CGM.
Since OH24 did not treat grain size distribution, we modify their framework.
The evolution of grain size distribution by various processes is
formulated based on our previous models, especially taken from \citet{Asano:2013aa} and
\citet{Hirashita:2019aa}.
Since dust enrichment is tightly related to metal production
(e.g., \cite{Lisenfeld:1998aa}), we model the chemical evolution in the CGM and the galaxy.
We treat each of the CGM and the central galaxy as a one-zone object, of which the spatial
structure is neglected.
{We treat the galaxy--CGM system as an isolated system and neglect the matter
circulation outside it.}

{Because we neglect detailed spatial structures in the galaxy and the CGM,
we assume that the grain size distribution in the outflow (inflow) is the same as that in the galaxy
(CGM). Indeed, a significant fraction of large grains, which are the major component of the
dust population in the galaxy, survive against sputtering in galactic outflows
\citep{Richie:2024aa}. Since we are not able to distinguish the gas once launched to the
CGM and that preexisting in the CGM, `dust destruction in the CGM' in our framework
includes sputtering both in the outflow and in the CGM.}

\subsection{Chemical evolution}\label{subsec:chem}

We calculate the enrichment of metals and dust in
the CGM and the galaxy. We describe the evolution of the gas, metal, and dust masses
in the galaxy, denoted as $M_\mathrm{g}$, $M_Z$, and $M_\mathrm{d}$,
respectively. The masses of the same components in the CGM are denoted with
superscript `C' (i.e., $M_\mathrm{g}^\mathrm{C}$, $M_Z^\mathrm{C}$, and
$M_\mathrm{d}^\mathrm{C}$). In our definition, the metals include both gas and dust phases.
The following equations are based on OH24, but are modified to include the grain size distribution.

In the galaxy, the evolution of the above mass components is described by
(e.g., \cite{Inoue:2011aa}):
\begin{eqnarray}
\frac{\mathrm{d}M_\mathrm{g}}{\mathrm{d}t} &=& -\psi +R+I-O,\label{eq:gas}\\
\frac{\mathrm{d}M_Z}{\mathrm{d}t} &=& -Z\psi +Y_Z+Z^\mathrm{C}I-ZO,\\
\frac{\mathrm{d}M_\mathrm{d}}{\mathrm{d}t} &=& -\mathcal{D}\psi +Y_\mathrm{d}
+\dot{M}_\mathrm{d,ISM}
+\mathcal{D}^\mathrm{C}I-\mathcal{D}O,
\end{eqnarray}
where $\psi$ is the star formation rate, $R$ is the gas return rate from stars,
$I$ is the inflow rate from the CGM, $O$ is the outflow rate to the CGM,
$Z\equiv M_Z/M_\mathrm{g}$ is the metallicity of the gas in the galaxy,
$Y_Z$ is the ejection rate of metals from stars,
$Z^\mathrm{C}\equiv M_Z^\mathrm{C}/M_\mathrm{g}^\mathrm{C}$ is the metallicity in the CGM,
$\mathcal{D}\equiv M_\mathrm{d}/M_\mathrm{g}$ is the
dust-to-gas ratio in the galaxy,
$\mathcal{D}^\mathrm{C}\equiv M_\mathrm{d}^\mathrm{C}/M_\mathrm{g}^\mathrm{C}$
is the dust-to-gas ratio in the CGM, and $\dot{M}_\mathrm{d,ISM}$ is the contribution to the
dust mass increase from interstellar dust processing (dust growth by accretion and dust
destruction by SN shocks).
Using the above three equations, we obtain the time evolution of $Z=M_Z/M_\mathrm{g}$
and $\mathcal{D}=M_\mathrm{d}/M_\mathrm{g}$ as
\begin{eqnarray}
\frac{\mathrm{d}Z}{\mathrm{d}t} &=& \frac{1}{M_\mathrm{g}}\left[ Y_Z-ZR+(Z^\mathrm{C}-Z)I\right] ,
\label{eq:metallicity}\\
\frac{\mathrm{d}\mathcal{D}}{\mathrm{d}t} &= & \frac{1}{M_\mathrm{g}}\left[ (Y_\mathrm{d}
-\mathcal{D}R)+(\mathcal{D}^\mathrm{C}-\mathcal{D})I+\dot{M}_\mathrm{d,ISM}\right] .\label{eq:dg}
\end{eqnarray}
We use equation (\ref{eq:metallicity}) to calculate the metallicity, but we do \textit{not} use
equation (\ref{eq:dg}) since interstellar processing is directly evaluated through the evolution of
grain size distribution formulated in subsection \ref{subsec:gsd}.
Instead, we calculate the dust-to-gas ratio contributed from stars (denoted
as $\mathcal{D}_\star$) by evaluating the stellar terms
in equation (\ref{eq:dg}):
\begin{eqnarray}
\frac{\mathrm{d}\mathcal{D}_\star}{\mathrm{d}t}=(Y_\mathrm{d}-\mathcal{D}_\star R)/M_\mathrm{g}.
\label{eq:dg_star}
\end{eqnarray}
A similar approach is taken by \citet{Hirashita:2020aa}.

In the CGM, the evolution is described by
\begin{eqnarray}
\frac{\mathrm{d}M_\mathrm{g}^\mathrm{C}}{\mathrm{d}t} &=& O-I,\label{eq:gasC}\\
\frac{\mathrm{d}M_Z^\mathrm{C}}{\mathrm{d}t} &=& ZO-Z^\mathrm{C}I,\\
\frac{\mathrm{d}M_\mathrm{d}^\mathrm{C}}{\mathrm{d}t} &=&
\mathcal{D}O-\mathcal{D}^\mathrm{C}I+
\dot{M}_\mathrm{d,CGM},
\end{eqnarray}
where $\dot{M}_\mathrm{d,CGM}$ is the changing rate of dust mass
by dust processing in the CGM.
From the above three equations, we obtain the time evolution of
$Z^\mathrm{C}=M_Z^\mathrm{C}/M_\mathrm{g}^\mathrm{C}$ and
$\mathcal{D}^\mathrm{C}=M_\mathrm{d}^\mathrm{C}/M_\mathrm{g}^\mathrm{C}$
as
\begin{eqnarray}
\frac{\mathrm{d}Z^\mathrm{C}}{\mathrm{d}t} &=& \frac{1}{M_\mathrm{g}^\mathrm{C}}
(Z-Z^\mathrm{C})O,\label{eq:metallicityC}\\
\frac{\mathrm{d}\mathcal{D}^\mathrm{C}}{\mathrm{d}t} &=& \frac{1}{M_\mathrm{g}^\mathrm{C}}
\left[ (\mathcal{D}-\mathcal{D}^\mathrm{C})O+\dot{M}_\mathrm{d,CGM}\right] .\label{eq:dgC}
\end{eqnarray}
We use equation~(\ref{eq:metallicityC}) to calculate the metallicity evolution in the CGM,
while we do \textit{not} solve equation (\ref{eq:dgC}) but directly use the dust evolution model in
subsection \ref{subsec:gsd}. We still refer to equation (\ref{eq:dgC})
together with equation (\ref{eq:dg}) later in formulating the effect of
mass exchange between the galaxy and the CGM (sub-subsection \ref{subsubsec:exchange}).

In summary, we use equations (\ref{eq:gas}) and (\ref{eq:metallicity}) to obtain
the evolution of the gas and metal content in the galaxy
and apply equations (\ref{eq:gasC}) and (\ref{eq:metallicityC})
to calculate that in the CGM.
For the dust evolution, we directly solve the equations formulated later in
subsection~\ref{subsec:gsd}, but
we use equation (\ref{eq:dg_star}) to evaluate the contribution
from stellar dust sources
(sub-subsection \ref{subsubsec:gsd_gal}).
We do not use equations (\ref{eq:dg}) and (\ref{eq:dgC}), but we
refer to these equations in evaluating the effects of inflow and outflow on the grain size distribution
(sub-subsection \ref{subsubsec:exchange}).

The rates of star formation, inflow, and outflow are determined by time-scales
($\tau_\mathrm{SF}$, $\tau_\mathrm{in}$, and $\tau_\mathrm{out}$, respectively),
which are treated as constant free parameters:
$\psi (t)=M_\mathrm{g}(t)/\tau_\mathrm{SF}$,
$I(t)=M_\mathrm{g}^\mathrm{C}(t)/\tau_\mathrm{in}$, and
$O(t)=M_\mathrm{g}/\tau_\mathrm{out}$.
We expect that the inflow and outflow are related to the star formation activity
in the galaxy. Thus, we assume $\tau_\mathrm{in}=\alpha\tau_\mathrm{SF}$
and $\tau_\mathrm{out}=\beta\tau_\mathrm{SF}$ and treat $\alpha$ and $\beta$ as free
parameters.
{Instead of determining these parameters by conducting detailed hydrodynamical
calculations, we move them in a wide range. This serves to investigate the effects of
each process on the resulting grain size distribution.}
{Also, the above parameterization assumes smooth changes in the SFR,
and inflow/outflow rates. \citet{Yajima:2017aa} showed using cosmological hydrodynamical
simulations that star formation may be intermittent at the beginning of galaxy evolution when
the galaxy mass is low but that it is smooth at low redshift. Moreover, we are interested in
evolution on a Gyr time-scale, which is much longer than the intermittence. Therefore,
the smooth SFR realized in our analytic work is reasonably applicable to this paper.}
Based on the star formation history, the mass ejection rates from stars
($R$, $Y_Z$ and $Y_\mathrm{d}$ for gas, metals, and dust, respectively) are
calculated by equations (8)--(10) of \citet{Hirashita:2020aa}
using stellar yield calculation data quoted in the same paper.

\subsection{Evolution of grain size distribution}\label{subsec:gsd}

The evolution of grain size distribution is calculated based on our previous
models, especially those developed by \citet{Hirashita:2019aa} and \citet{Hirashita:2021ab}.
The grain size distribution at time $t$ is defined as the number density of dust grains
in the radius range between $a$ and $a+\mathrm{d}a$. We calculate the grain
size distributions in the galaxy and in the CGM, which are denoted as
$n(a,\, t)$ and $n^\mathrm{C}(a,\, t)$, respectively. In equations that
describe the evolution of grain size distribution,
we use the grain mass distribution in the galaxy
\begin{eqnarray}
\rho_\mathrm{d}(m,\, t)\,\mathrm{d}m=\frac{4}{3}\pi a^3sn(a,\, t)\,\mathrm{d}a,
\label{eq:rho_n}
\end{eqnarray}
where the grain mass is calculated by assuming a spherical shape as $m=(4\pi /3)a^3s$
($s$ is the material density).
We also define $\tilde{\rho}_\mathrm{d}$ as the grain mass distribution normalized to the
gas mass density: $\tilde{\rho}_\mathrm{d}\equiv\rho_\mathrm{d}/(\mu m_\mathrm{H}n_\mathrm{H})$,
where $\mu$ is the gas mass per hydrogen mass, and
$m_\mathrm{H}$ is the hydrogen atom mass.
The grain mass distribution is linked to the dust-to-gas ratio as
\begin{eqnarray}
\mathcal{D}(t)=\int_0^\infty\tilde{\rho}_\mathrm{d}(m,\, t)\,\mathrm{d}m.
\end{eqnarray}
The grain mass distribution in the CGM is defined in the same way and denoted with
superscript `C' as
$\rho_\mathrm{d}^\mathrm{C}(m,\, t)$ and $\tilde{\rho}_\mathrm{d}^\mathrm{C}(m,\, t)$.

The grain size distribution is discretized into 128 logarithmically spaced grain radius bins
in the range of $a=0.01$--10 $\micron$ following appendix B of \citet{Hirashita:2019aa}.
For the boundary condition, we
adopt $n=0$ and $n^\mathrm{C}=0$
(or equivalently $\rho_\mathrm{d}=\tilde{\rho}_\mathrm{d}=0$ and
$\rho_\mathrm{d}^\mathrm{C}=\tilde{\rho}_\mathrm{d}^\mathrm{C}=0$)
at the maximum and minimum grain radii.

\subsubsection{In the galaxy}\label{subsubsec:gsd_gal}

We calculate the evolution of grain size distribution in the galaxy based on
\citet{Hirashita:2019aa} and \citet{Hirashita:2020aa}.
We refer the interested reader to these references
for detailed equations. In what follows, we review the calculation of grain size distribution in
each process and describe the changes made to include the mass exchange between the
galaxy and the CGM.

We include stellar dust production, shattering in the diffuse ISM,
dust growth by the accretion of gas-phase metals in the dense gas (this process is referred to
as accretion),
coagulation in the dense ISM, and destruction in SN shocks (referred to as SN destruction).
For simplicity, we separate the ISM into dense and diffuse phases and assume
the mass fraction of the dense ISM to be 0.5 (i.e., the other half is occupied by the diffuse ISM).
The effect of varying the dense gas fraction has already been examined by
\citet{Hirashita:2020aa}, and the adopted fraction 0.5 explains well the Milky Way extinction curve.
The time-step is weighted with the dense gas fraction (0.5) for accretion and coagulation,
and with the diffuse gas fraction (0.5) for shattering.
The densities and temperatures of the dense and diffuse phases are
$(n_\mathrm{H}/\mathrm{cm}^3,\, T_\mathrm{gas}/\mathrm{K})=(300,\, 25)$
and (0.3, $10^4$), respectively.
For the other processes (stellar dust production and SN destruction), we use the full time-step.

For stellar dust production, the increasing rate of grain size distribution
at radius $a$ (or equivalently at mass $m$) is evaluated using
$\mathrm{d}\mathcal{D}_\star /\mathrm{d}t$ (equation \ref{eq:dg_star}).
We distribute this newly supplied dust
into grain radius bins by assuming a lognormal distribution centered at $a=0.1~\micron$
with a standard deviation of 0.47 (see also \cite{Asano:2013aa}).

For accretion and SN destruction, we adopt the changing rate of $m$ ($\dot{m}$, where
the dot indicates the time derivative) that
depends on the grain radius. As a consequence, the grain size distribution is governed
by advection equations with appropriate $\dot{m}$ (or $\dot{a}$).
The accretion rate is estimated by the collision rate between grains and
gas-phase metals.
Following \citet{Hirashita:2023aa}, we set a maximum for the dust-to-metal ratio,
$\mathrm{(D/M)_{max}}=0.48$; that is, accretion is turned off if the dust-to-metal ratio exceeds
this value. This maximum value is
adopted to match the dust extinction and emission per hydrogen in the Milky Way,
and is also based on the consideration that
some metal elements are not easily incorporated into the dust (solid) phase.
For SN destruction, we estimate the dust destruction rate based on the sweeping rate of
SN shocks in the ISM (the SN rate is evaluated in a manner consistent
with the star formation history, $\psi (t)$)
with an assumed efficiency of destruction as a function grain radius, $\epsilon_\mathrm{dest}(a)$:
\begin{eqnarray}
\epsilon_\mathrm{dest}(a)=1-\exp\left[-0.1(a/0.1~\micron )^{-1}\right] .\label{eq:epsilon}
\end{eqnarray}

For coagulation and shattering, we solve Smoluchowski equations using a Kernel function
evaluated with geometrical cross-sections and grain velocities.
We consider the grain--gas coupling in
a Kolmogorov turbulence to obtain the typical grain velocity as a function of grain radius.
For shattering, we adopt the total fragment mass formed in a collision and the fragment
size distribution function from \citet{Kobayashi:2010aa}.

For coagulation, we newly include the coagulation threshold velocity,
above which grain--grain collisions do not lead to sticking. This is to avoid run-away
coagulation, which could occur because larger grains have higher velocities.
In reality, grains should bounce or fragmented in high-velocity collisions.
Thus, we multiply the Kernel function ($\mathcal{K}$)\footnote{In \citet{Hirashita:2019aa},
the Kernel function is denoted as $\alpha$. Since $\alpha$ is used to
parameterize the inflow time-scale in this paper (subsection~\ref{subsec:chem}),
we adopt a different notation.} used in \citet{Hirashita:2019aa} by
an exponential cut-off function for coagulation:
\begin{eqnarray}
\mathcal{K} (m_1,\, m_2)=\frac{\sigma_{1,2}v_{1,2}}{m_1m_2}
\exp\left[-\left(\frac{v_{1,2}}{v_\mathrm{th,coag}}\right)^2\right]\, ,
\end{eqnarray}
where $m_1$ and $m_2$ are the masses of the two colliding grains,
$v_{1,2}$ is the relative velocity (with a random direction),
$\sigma_{1,2}=\pi (a_1+a_2)^2$
is the cross-section ($a_1$ and $a_2$ are the grain radii of the two grains),
and $v_\mathrm{th,coag}$ is the coagulation threshold velocity above which
grains do not stick.
We adopt $v_\mathrm{th,coag}=0.1$ km~s$^{-1}$, which is near to the velocity
($\sim 0.08$ km~s$^{-1}$) above which
collisions could disrupt grains \citep{Wada:2013aa}. As argued by
\citet{Hirashita:2014ab}, lower thresholds suggested experimentally
by \citet{Blum:2000aa} and theoretically by \citet{Dominik:1997aa}
fail to produce grains larger than 0.1~$\micron$. Indeed, the above coagulation threshold
velocity reproduces the maximum grain radius ($\simeq 0.25~\micron$; \cite{Mathis:1977aa},
hereafter MRN) for the Milky Way.

\subsubsection{In the CGM}\label{subsubsec:gsd_cgm}

The evolution of grain size distribution in the CGM is driven by
sputtering and shattering. Sputtering takes place in the hot phase with
$T_\mathrm{gas}\gtrsim 10^6$ K (e.g., \cite{Draine:1979aa,Tsai:1995aa,Hirashita:2015aa}),
while shattering occurs
in a colder medium (referred to as cool clumps; \cite{Hirashita:2021ab}).
For simplicity, we assume the weighting factor (or the mass fraction)
for these two phases to be 0.5;
that is, a single time-step $\Delta t$ is divided into two steps ($0.5\Delta t$ for
each) and we treat shattering and sputtering in each of these time-steps.
These weighting factors are degenerate with the efficiencies of shattering and sputtering.
Thus, we fix the mass fractions of the two phases and move the efficiencies of the relevant
processes as we will explain below.

The equation describing sputtering is the same as that used for SN destruction in the galaxy,
but we use a different destruction rate.
We adopt the following formula for the time-scale of dust destruction by sputtering,
utilizing $\epsilon_\mathrm{dest}(a)$ in equation (\ref{eq:epsilon}):
\begin{eqnarray}
\dot{m}=-\epsilon_\mathrm{dest}(a)m/\tau_\mathrm{dest},
\end{eqnarray}
where $\tau_\mathrm{dest}$ is the destruction time-scale treated as a free parameter.
Note that
$m$ and $a$ are related as described below equation (\ref{eq:rho_n}).
In the hot phase, sputtering occurs even more quickly than the
exchange between the cool and hot phases.
Small grains formed in cool clumps by shattering do not suffer sputtering before they are injected
into the hot medium.
Thus, we effectively regard $\tau_\mathrm{dest}$ as the phase exchange time-scale between the
cool and hot gas. We incorporate the above form of $\dot{m}$ into
the equation describing the evolution of
grain size distribution by sputtering
(see equation 7 of \cite{Hirashita:2019aa}).

For shattering, we use the method developed by \citet{Hirashita:2021ab}.
We adopt their fiducial values for the gas density and temperature in cool clumps
($n_\mathrm{H}=0.1$~cm$^{-1}$ and $T_\mathrm{gas}=10^4$~K).
The grain velocity, which contributes to the
kernel function in the shattering equation, is determined by the turbulence model described in
the same paper. The turbulence is assumed to have a
Kolmogorov spectrum with injection scale $L_\mathrm{max}$ and maximum velocity
$v_\mathrm{max}$. We adopt the fiducial values for these two parameters
($L_\mathrm{max}=100$ pc and $v_\mathrm{max}=10$ km s$^{-1}$).
Larger grains, coupled with larger eddies, tend to have higher velocities.
At the same time, if grains are too large (typically $a\gtrsim 0.1~\micron$), they are not coupled
even with the largest eddies. Therefore, the grain velocity
becomes a decreasing function of $a$ at $a\gtrsim 0.1~\micron$.
Based on this grain velocity model, we solve the
same shattering equation as used for the dust in the galaxy.

Since $n_\mathrm{H}$ predominantly affects the shattering efficiency,
we specifically denote $n_\mathrm{H}$ in cool clumps as $n_\mathrm{H,cool}^\mathrm{C}$,
and vary it.
The gas density in cool clumps affects the grain size distribution in the CGM
in the following two aspects \citep{Hirashita:2021ab}:
(i) The grain--grain collision rate is enhanced in denser (larger $n_\mathrm{H,cool}^\mathrm{C}$)
environments.
(ii) Large grains (typically at $a\gtrsim 0.1~\micron$) attain higher velocities
for higher $n_\mathrm{H,cool}^\mathrm{C}$ because
the grains are more strongly coupled with the gas (turbulence). Because of these two effects,
the shattering rate in the CGM is sensitive to $n_\mathrm{H,cool}^\mathrm{C}$.

\subsubsection{Effect of the CGM--galaxy mass exchange}\label{subsubsec:exchange}

In addition to the dust processing in the galaxy and CGM, we also need to include the
mass exchange between these two zones through the outflow and inflow.
The inflow changes the grain mass distribution in the galaxy as
\begin{eqnarray}
\left[\frac{\partial\tilde{\rho}_\mathrm{d}}{\partial t}\right]_\mathrm{inflow}=
\frac{1}{M_\mathrm{g}}\left(\tilde{\rho}_\mathrm{d}^\mathrm{C}-\tilde{\rho}_\mathrm{d}\right) I.
\end{eqnarray}
The inflow does not affect the grain mass distribution in the CGM.
The outflow causes the following change for the grain mass distribution in the CGM:
\begin{eqnarray}
\left[\frac{\partial\tilde{\rho}_\mathrm{d}^\mathrm{C}}{\partial t}\right]_\mathrm{outflow} =
\frac{1}{M_\mathrm{g}^\mathrm{C}}
\left(\tilde{\rho}_\mathrm{d}-\tilde{\rho}_\mathrm{d}^\mathrm{C}\right) O.
\end{eqnarray}
The outflow does not affect the grain mass distribution in the galaxy.
These terms are similar to the inflow and outflow terms in the equations for
$\mathcal{D}$ and $\mathcal{D}^\mathrm{C}$ (equations \ref{eq:dg} and \ref{eq:dgC},
respectively).
Indeed, if we integrate the above two equations for $m$, we obtain the corresponding
terms for equations (\ref{eq:dg}) and (\ref{eq:dgC}).
The inflow and outflow tend to make the grain size distributions in the CGM and in the galaxy
approach each other on timescales $M_\mathrm{g}/I$ and $M_\mathrm{g}^\mathrm{C}/O$,
respectively.

\subsection{Settings and parameters}\label{subsec:param}

We start calculation from $M_\mathrm{g}=0$ and $M_\mathrm{g}^\mathrm{C}=M_0$
with a metal (dust)-free condition
($Z=Z^\mathrm{C}=0$ and $\rho_\mathrm{d}=\rho_\mathrm{d}^\mathrm{C}=0$)
at $t=0$. In this situation, the galaxy builds up its baryonic content by the inflow.
We adopt $M_0=10^{11}$~M$_{\odot}$ as a tentative value,
aiming roughly at the baryonic mass in the Milky Way.
This is the same as the CGM mass in OH24.
While the CGM gas mass is constant in OH24 (i.e., the CGM is treated as a gas
reservoir), it is decreased by the infall in our model.
Thus, we effectively adopt a smaller mass scale
than OH24.
However, {since we output the grain size distributions per gas mass
(subsection \ref{subsec:gsd}), it} is
independent of the total mass scale. Therefore,
the choice of $M_0$ {does not directly influence} the conclusions in this paper.

For the grain material density,
we adopt $s=3.5$ g cm$^{-2}$, which is taken from astronomical silicate in
\citet{Weingartner:2001aa}.
Carbonaceous dust has a lower density; thus, with a fixed dust abundance, the grain
number density is larger. This leads to somewhat faster interstellar processing such as
small grain formation by shattering. Carbonaceous dust also has small tensile strength,
so that it is more easily shattered \citep{Jones:1996aa}.
Therefore, adopting silicate properties would serve to examine a conservative case
for the effectiveness of interstellar processing.
However, the difference in grain species affects the grain size distribution less than
the variations in the efficiencies of essential physical processes;
that is, the differences in $\tau_\mathrm{in}$, $\tau_\mathrm{out}$,
$\tau_\mathrm{sput}$, and $n_\mathrm{H,cool}^\mathrm{C}$ are more significant.
Thus, we focus on these parameters, fixing the dust material properties.

As described in subsection~\ref{subsec:chem}, the inflow and outflow are regulated by
the parameters $\alpha$ and $\beta$, respectively. We vary
$\alpha$ and $\beta$ with $\tau_\mathrm{SF}=5$~Gyr fixed to the fiducial value adopted
by OT24 (see also \cite{Asano:2013aa,Hirashita:2020aa}).
By fixing $\tau_\mathrm{SF}$, we concentrate on the variations of $\alpha$ and $\beta$.
We expect that $\alpha$ and $\beta$ are not much above or below unity since
otherwise, the gas masses of the galaxy and the CGM are strongly imbalanced.\footnote{To
avoid this imbalance, OH24
assumed a constant CGM mass (i.e.\ the CGM acted as a gas reservoir in their model)
as mentioned above.
However, this assumption is not essential in the parameter ranges of interest.}
Thus, we adopt $\alpha =1$ and $\beta =1$ for the fiducial values, and
investigate the range of $\alpha =0.1$--10 and $\beta =0.1$--10.
We note that $1/\beta$ corresponds to the mass-loading factor $\eta_\mathrm{out}$ in OT24.
Some hydrodynamical simulations indicate that $\beta\sim 1$ (or $\eta_\mathrm{out}\sim 1$;
e.g., \cite{Hu:2023aa}), although it is sensitive to
the treatment of stellar feedback and is dependent on the galactocentric distance.
{We do not directly include AGN feedback.
\citet{Valiante:2011aa} included galactic outflows driven by both SNe and AGNs in the dust
evolution model based on their semi-analytic model.
Because of the co-evolution of stellar and black-hole masses as is also reproduced by
their model,
we expect that AGN feedback is indirectly related to the stellar feedback.
Thus, we could
regard $\beta$ as effectively including AGN-driven outflows, although
we need to develop a model that includse the evolution of the central black hole
and the CGM at the same time (e.g., \cite{Ramesh:2023aa}) to isolate the effect of AGNs.}

The sputtering time-scale $\tau_\mathrm{sput}$ is related to
the exchange time-scale between the cool and hot gas phases
(sub-subsection~\ref{subsubsec:gsd_cgm}).
OH24 adopted a sputtering time-scale of 2~Gyr based on the gradual decline of
the dust mass density in the CGM seen in a hydrodynamic simulation \citep{Aoyama:2018aa}.
However, OH24 did not consider the grain radius dependence,
and we expect that the time-scale is shorter for small grains whose destruction efficiency is
almost unity ($\epsilon_\mathrm{dest}\sim 1$).
In this paper, we determine $\tau_\mathrm{dest}$ in the following way.
The lifetime of cool clumps can be as long as 0.1--1 Gyr based on analytic estimates
\citep{Lan:2019aa} and hydrodynamical simulations \citep{Armillotta:2017aa}, but could be
uncertain because of the details in thermal conduction and magnetic field
\citep{Li:2020aa,Sparre:2020aa}. We adopt $\tau_\mathrm{sput}=0.3$~Gyr as a
fiducial value and examine a range of 0.03--3 Gyr, considering the uncertainties.
As mentioned in sub-subsection~\ref{subsubsec:gsd_cgm}, the gas density in cool clumps
has a large influence on the shattering efficiency. We vary $n_\mathrm{H,cool}^\mathrm{C}$
in the range of 0.03--0.3 cm$^{-3}$ with a fiducial value of 0.1~cm$^{-3}$
\citep{Hirashita:2021ab}. The variations of these parameters serve to clarify how sputtering and shattering
affect the grain size distribution in the CGM.

The varied parameters are summarized in table \ref{tab:param}, where we show
the fiducial values and the ranges. Since we model grain processing in the
grain-size-dependent way, the parameterization is not exactly the same as that in OT24.
Thus, we do not fine-tune the parameters but investigate a wide range for each of them.
This approach is also useful to investigate
the response of the grain size distribution to the relevant parameters.

\begin{table}
\tbl{Parameters.}{%
\begin{tabular}{lcccc}
\hline
Parameter & units & fiducial value & minimum & maximum \\
\hline
$\alpha$ & & 1 & 0.1 & 10 \\
$\beta$ & & 1 & 0.1 & 10\\
$\tau_\mathrm{sput}$ & Gyr & $0.3$ & $0.03$ & $3$\\
$n_\mathrm{H,cool}^\mathrm{C}$ & cm$^{-3}$ & 0.1 & 0.01 & 1 \\
\hline
\end{tabular}}
\label{tab:param}
\end{table}

\section{Results}\label{sec:result}

\subsection{Dust and gas masses in the CGM}\label{subsec:mass}

Although our main focus is laid on the grain size distribution, it is useful to
confirm if our model is consistent with the evolution of the total dust mass shown in
our previous paper (OH24). The dust mass can be `decomposed' into the gas mass
and the dust-to-gas ratio, which are also examined here. In addition, we
also present the dust-to-metal ratio to show the efficiency of dust production out of metals.

\begin{figure}
 \begin{center}
 \includegraphics[width=0.9\columnwidth]{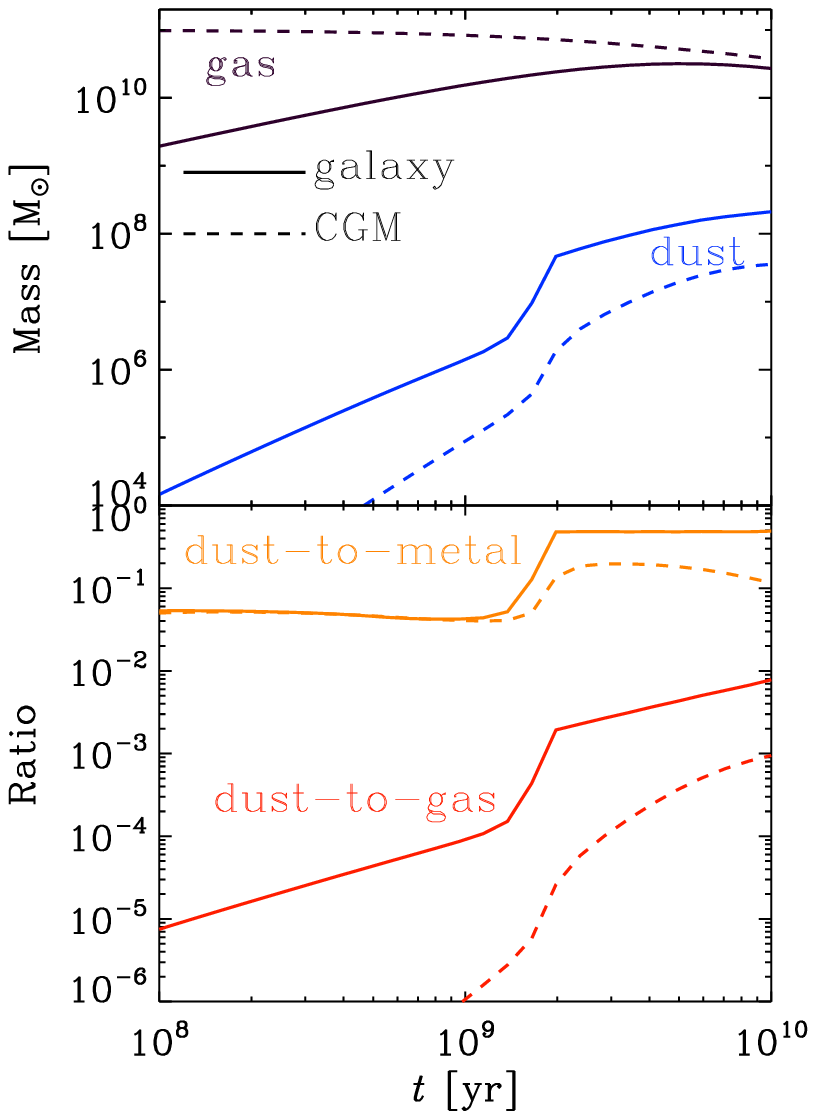}
 \end{center}
 \caption{Evolution of the gas (black lines) and dust (blue lines) masses
 (upper window) and
 the dust-to-gas (red) and dust-to-metal (orange) ratios (lower window).
 The solid and dashed lines show the quantities in the galaxy and in the CGM,
 respectively.}
 \label{fig:mass_std}
\end{figure}

In figure \ref{fig:mass_std}, we show the evolution of the quantities related to the total dust mass.
The evolution of the gas mass directly traces the mass exchange between
the CGM and the galaxy. The gas mass in the CGM decreases because of the infall into the
galaxy although there is a partial return from the galaxy through the outflow.
The gas mass in the galaxy grows in the early epochs, but it decreases later because
the gas consumption by star formation becomes more prominent than the gas supply from the
CGM. The dust masses in both the galaxy and the CGM rise, which is also consistent with the
increase of the dust-to-gas ratio indicated in the lower window. The steep increases of
the dust mass and the dust-to-gas ratio at $t\sim 2$ Gyr are due to dust growth by accretion,
as commonly seen in previous dust evolution models
(e.g., \cite{Dwek:1998aa,Hirashita:1999aa}).
Although this growth only occurs in the galaxy, it affects the dust abundance in the CGM indirectly
through the dust transport by the outflow.
The dust-to-metal ratio behaves like a
step function: It is determined by the dust
condensation efficiency in stellar ejecta at low metallicity, while it reaches the maximum value
(0.48; subsection \ref{subsec:gsd}) because of accretion at high metallicity.
At later stages, the dust-to-metal ratio in the CGM declines
because of sputtering.

\begin{figure}
 \begin{center}
 \includegraphics[width=0.49\columnwidth]{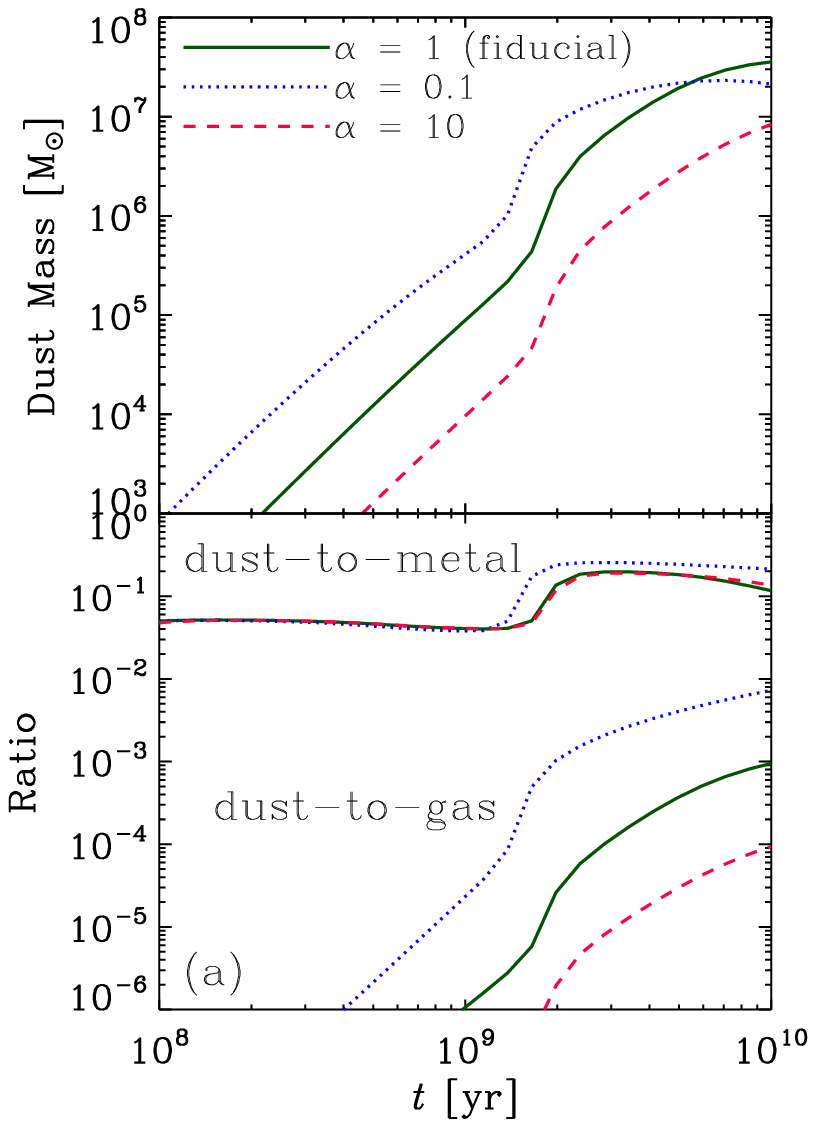}
 \includegraphics[width=0.49\columnwidth]{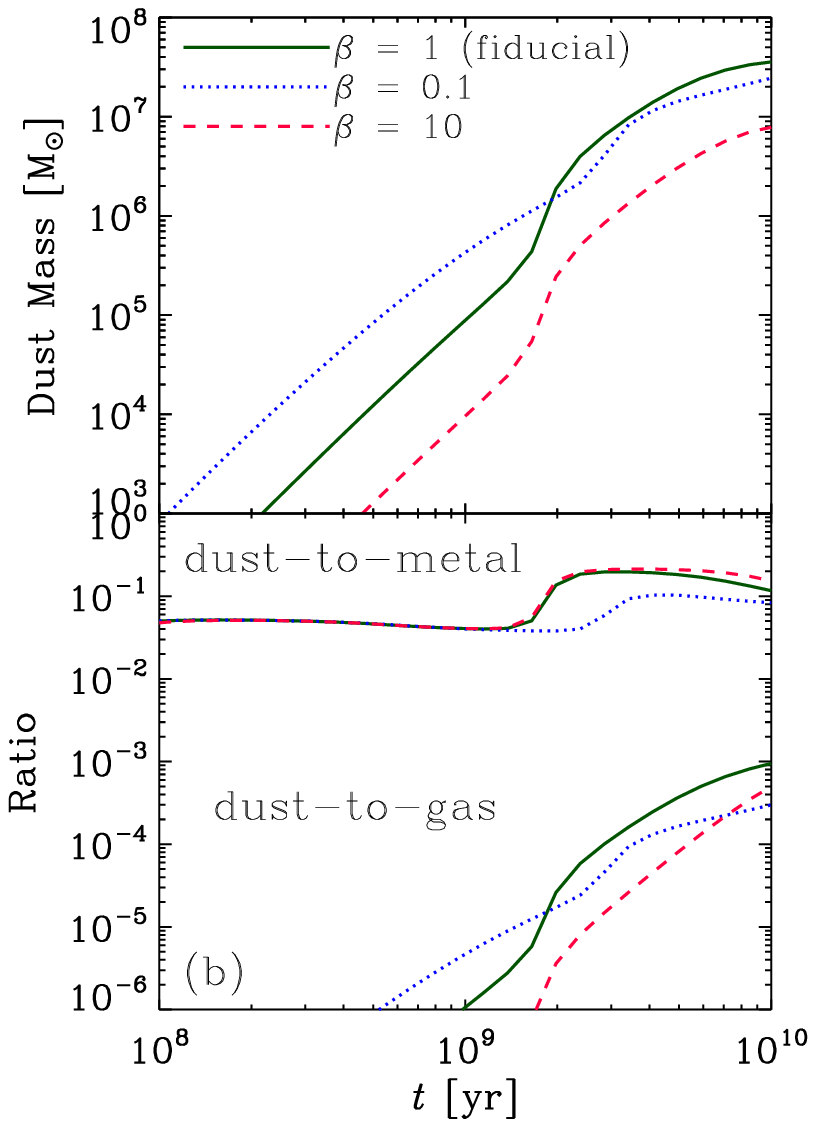}
 \end{center}
 \caption{Evolution of the dust mass in the CGM
 (upper window) and
 the dust-to-gas and dust-to-metal ratios in the CGM (lower window)
 for various values of the inflow and outflow parameters.
 Panels (a) and (b) show the variations of $\alpha$ and $\beta$,
 respectively. The efficiencies of inflow and outflow are inversely proportional to
 $\alpha$ and $\beta$, respectively.
 The correspondence between the value of $\alpha$ or $\beta$ and
 the line species is shown in the legend. We adopt the fiducial values for the parameters
 other than the varied one.}
 \label{fig:mass_inflow}
\end{figure}

We clarify the effects of galaxy--CGM mass exchange on the evolution
of dust content by varying $\alpha$ and $\beta$. We focus on dust-related quantities
in the CGM.
In figure~\ref{fig:mass_inflow}(a), we show the evolution of dust mass, dust-to-gas ratio,
and dust-to-metal ratio in the CGM for various values for $\alpha$.
The dust mass is higher for a smaller value of $\alpha$ in the early phase
($t\lesssim 3$ Gyr) because a higher star formation rate
caused by a more efficient inflow leads to quicker metal enrichment in both
the galaxy and the CGM. The increase of the dust mass is stopped at later epochs for
the case of $\alpha =0.1$,
which reflects the depletion of infalling gas.
The dust-to-gas and dust-to-metal ratios are higher for a smaller value of $\alpha$ because of
quicker metal and dust enrichment.
The smaller CGM mass in the case of smaller $\alpha$ also makes the dust enrichment
easier simply because there is less gas to enrich.
However, the effect of $\alpha$ on the dust-to-metal ratio is not as large as that on
the dust-to-gas ratio,
which indicates that the dust abundance is strongly correlated with the metallicity.

In figure \ref{fig:mass_inflow}(b), we show the effects of $\beta$, which regulates the outflow.
In the early epoch, the dust mass is almost inversely
proportional to $\beta$ since dust mass supply to the CGM is governed by the outflow rate
from the galaxy. At later times, the dust mass does not change monotonically with $\beta$:
The fiducial model achieves the largest dust mass among the three cases. If $\beta$ is
too small, the efficient outflow keeps the gas mass in the galaxy low, leading to inefficient dust
production. If $\beta$ is too large in contrast, the supply of dust from the galaxy
is inefficient so that the dust mass tends to be low in the CGM.
The evolution of the dust-to-gas ratio broadly follows that of the dust mass.
The dust-to-metal ratio in the CGM is smaller for smaller $\beta$, reflecting
a smaller dust-to-metal ratio (i.e., less efficient dust enrichment) in the galaxy.

\begin{figure}
 \begin{center}
 \includegraphics[width=0.49\columnwidth]{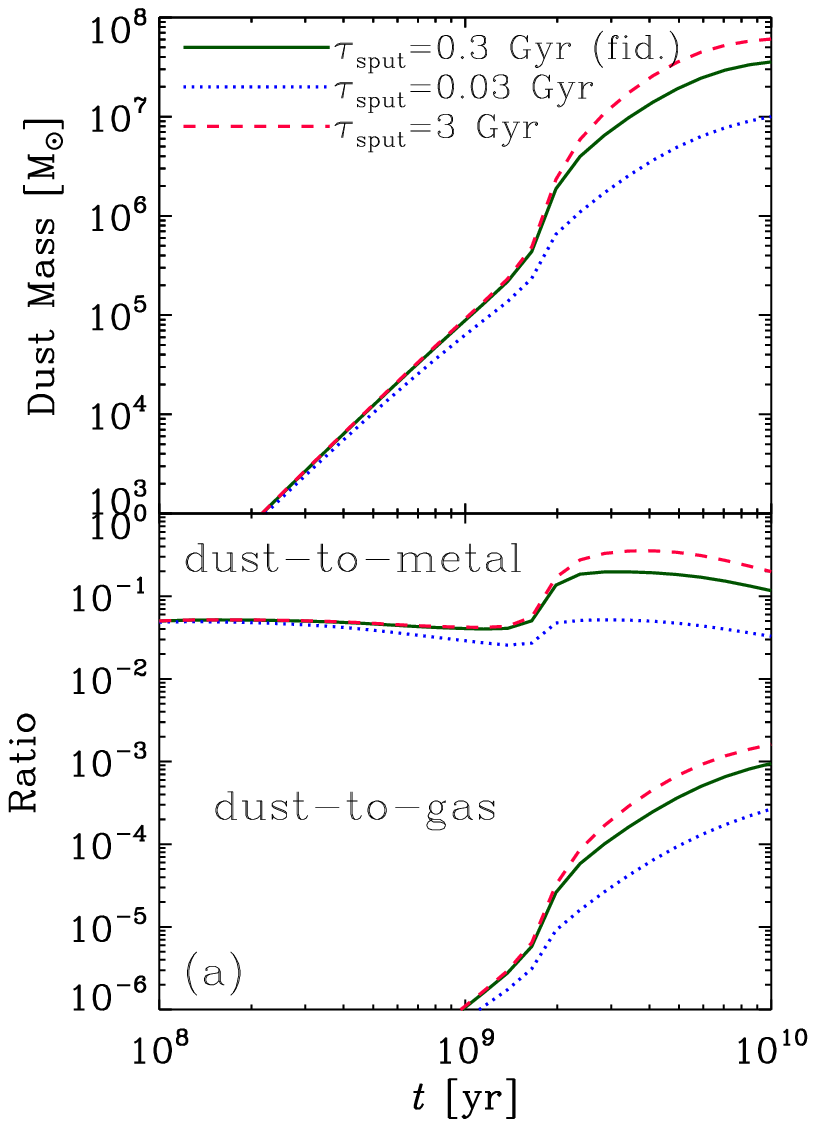}
 \includegraphics[width=0.49\columnwidth]{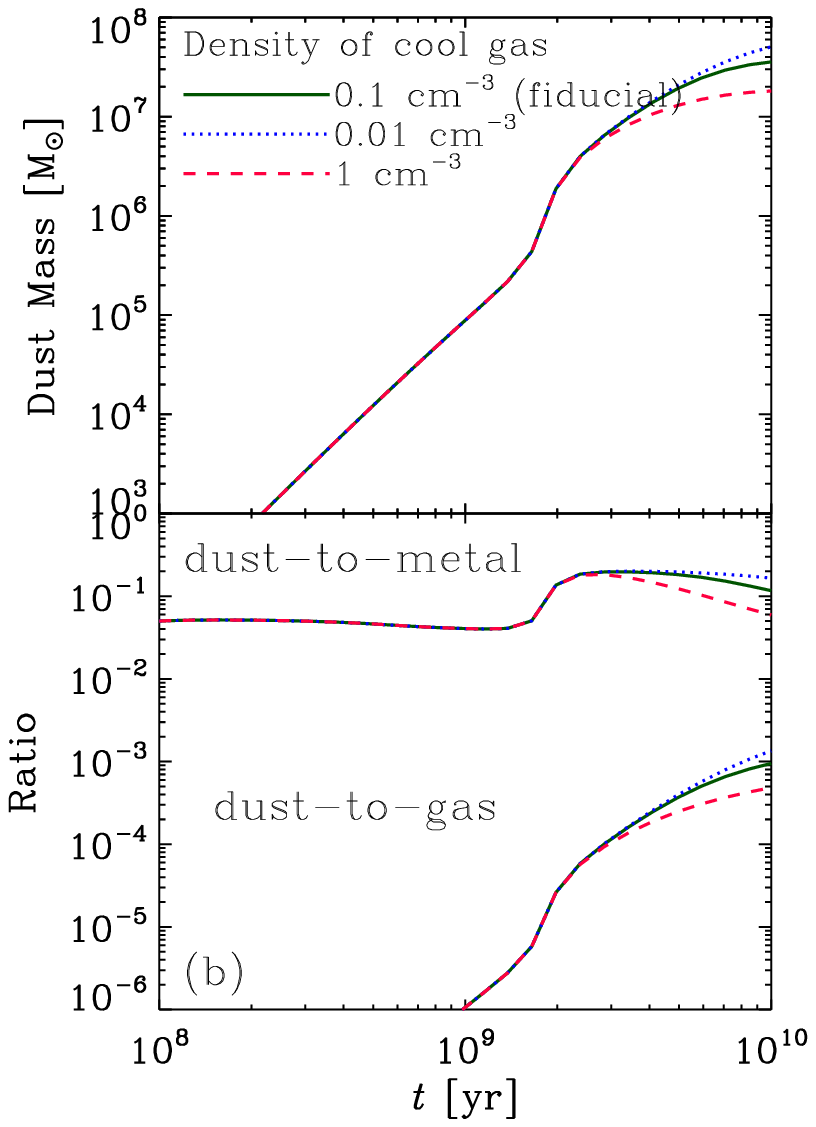}
 \end{center}
 \caption{Same as figure \ref{fig:mass_inflow} but
 for various values of the parameters that regulate sputtering and shattering in the CGM.
 Panels (a) and (b) show the variations of $\tau_\mathrm{sput}$ and
 $n_\mathrm{H,cool}^\mathrm{C}$, respectively. The correspondence between the value of
 these parameters and the line species is shown in the legend.
 We adopt the fiducial values for the parameters
 other than the varied one.}
 \label{fig:mass_sput}
\end{figure}

We also examine the effect of dust processing in the CGM, that is, sputtering and shattering.
In figure \ref{fig:mass_sput}(a), we show the evolution of dust-related quantities in the CGM
for various $\tau_\mathrm{sput}$ in order to examine the effect of sputtering.
We confirm that stronger sputtering reduces the dust mass and abundance in the CGM.
The effect of sputtering appears at an earlier epoch for shorter $\tau_\mathrm{sput}$.
Thus, we confirm that the dust destruction by sputtering has a direct impact on the
dust abundance in the CGM.

In figure \ref{fig:mass_sput}(b), we show the effects of shattering efficiency regulated
by $n_\mathrm{H,cool}^\mathrm{C}$. We observe that, if shattering is efficient with
large $n_\mathrm{H,cool}^\mathrm{C}$, the dust mass and abundance are slightly diminished.
Although shattering itself conserves the total dust mass, it makes the grain size distribution
biased to small grains, which are easily destroyed by sputtering.
Shattering, thus, has an indirect influence on the CGM dust abundance through
enhanced sputtering.
Since shattering becomes efficient after the CGM is enriched with a significant amount of dust,
the effect of shattering appears at later epochs.

It may be useful to compare our results with OH24's, although our models and parameterizations
are not the same as theirs.
In our fiducial model, the dust masses in the galaxy and in the CGM at $t=10$ Gyr are
$2\times 10^8$ and $4\times 10^7$ M$_{\odot}$, respectively;
in OH24's fiducial model, $3\times 10^8$ and $2\times 10^8$ M$_{\odot}$, respectively.
In our fiducial model, the stellar mass at $t=10$ Gyr is $4\times 10^{10}$ M$_{\odot}$, while
in OH24's, it is $6\times 10^{10}$ M$_{\odot}$.
As mentioned in subsection \ref{subsec:param}, OH24 assumed a constant gas mass for the
CGM, while we decrease it according to the inflow. Thus, we tend to underpredict
the stellar and metal/dust masses.

OH24 adopted the CGM dust mass from \citet{Menard:2010aa},
who showed from statistical analysis of background quasar colors
that the dust mass in the CGM is on average $5\times 10^7$ M$_{\odot}$
for a sample of $L^*$ galaxies whose mean redshift is $z\sim 0.36$.
{\citet{Peek:2015aa} obtained a similar dust mass in the CGM for a lower-redshift
sample.}
As estimated by OH24, the CGM dust mass normalized to the stellar mass
is $\sim 1\times 10^{-2}$, while it is
$\sim 1\times 10^{-3}$ in our model.
Thus, our dust enrichment model predicts an order-of-magnitude smaller
dust abundance in the CGM compared with OH24, whose model explained
the CGM dust abundance of the above observational sample.
However, our value is rather near to the one indicated by stacking analysis of
extended dust emission around galaxies at $z\sim 1$ \citep{Meinke:2023aa}.
Considering the uncertainties in the observational data in terms of the dust and
stellar masses, we do not further tune the fiducial model,
but we concentrate on the effects of various parameters
on the resulting grain size distribution.
We still discuss observational aspects including the above discrepancy later in
section \ref{sec:discussion}.


\subsection{Evolution of grain size distribution}\label{subsec:gsd_result}

\begin{figure}
 \includegraphics[width=\columnwidth]{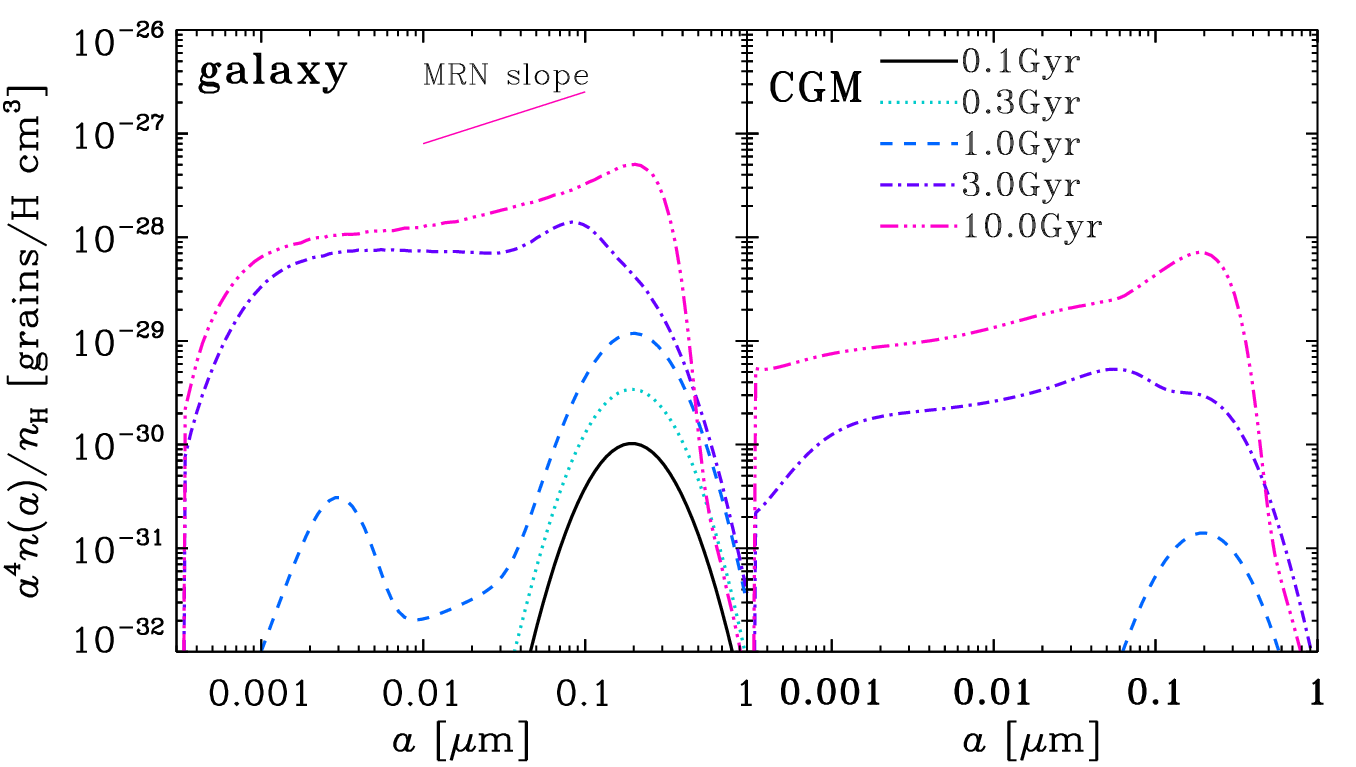}
 \caption{Evolution of grain size distribution in the galaxy (left panel) and in the CGM
 (right panel) for the fiducial model.
 The grain size distribution $n$ multiplied by $a^4/n_\mathrm{H}$ is proportional to
 the dust-to-gas ratio in $\log a$ bins.
 The grain size distributions are shown for ages
 $t=0.1$, 0.3, 1, 3, and 10~Gyr (solid, dotted, dashed, and dot--dashed, and
 triple-dot--dashed lines, respectively). In the CGM, the grain size distribution at
 $t=0.1$ Gyr does not emerge
 in this figure because the dust abundance is too low.}
 \label{fig:gsd}
\end{figure}

We first discuss the case with the fiducial parameter values (table~\ref{tab:param}).
We show the evolution of grain size distribution in the galaxy and in the CGM
in figure~\ref{fig:gsd}.

The evolution of grain size distribution (figure \ref{fig:gsd} left)
in the galaxy is similar to that presented in our previous models;
thus, we refer the reader to  \citet{Hirashita:2019aa} and \citet{Hirashita:2020aa}
for detailed discussions on the evolution of grain size distribution in galaxies.
We briefly describe the outline here.
Overall, the grain size distribution rises in almost all the grain radius range
as a function of time because of the increase in the grain abundance.
In the early epoch, dust production is dominated by stellar sources, which
mainly supply large ($a\sim 0.1~\micron$) grains.
After that, shattering becomes efficient enough to convert large grains to small grains,
whose abundance is further raised by accretion at $t\sim 1$~Gyr. The effect of accretion is
seen in the bump at $a\sim 0.003~\micron$.
At $t=\mbox{a few Gyr}$, the abundance of small grains is large enough to cause
efficient coagulation, which smoothes out the bump into a power-law like
grain size distribution. In the end, the grain size distribution approaches a shape similar to the
MRN distribution.

The grain size distribution in the CGM broadly follows that in the galaxy, which is
the main source of dust in the CGM. The grain size distribution is
dominated by large grains in the early epoch and is in a smooth power-law-like shape
later. The level of the grain size distribution in the CGM is lower than that in the galaxy,
reflecting the lower dust abundance and metallicity.

Although the slope of grain size distribution in the CGM is similar to that in the galaxy
in the fiducial model, it is not generally true.
A difference in the functional shape of grain size distribution between the
galaxy and the CGM is caused by sputtering and shattering in the CGM.
Sputtering destroys small grains more efficiently than large ones
while shattering increases the small grain abundance.
Thus, the balance between sputtering and shattering determines the slope of
the grain size distribution in the CGM. We further investigate
this balance in subsection \ref{subsec:sput_shat}.

\subsection{Effects of inflow and outflow}\label{subsec:inflow}

The mass exchange between the galaxy and the CGM drives the evolution of dust abundance
and grain size distribution in the CGM. We examine the effects of inflow and outflow by
varying $\alpha$ and $\beta$.

\begin{figure}
 \includegraphics[width=\columnwidth]{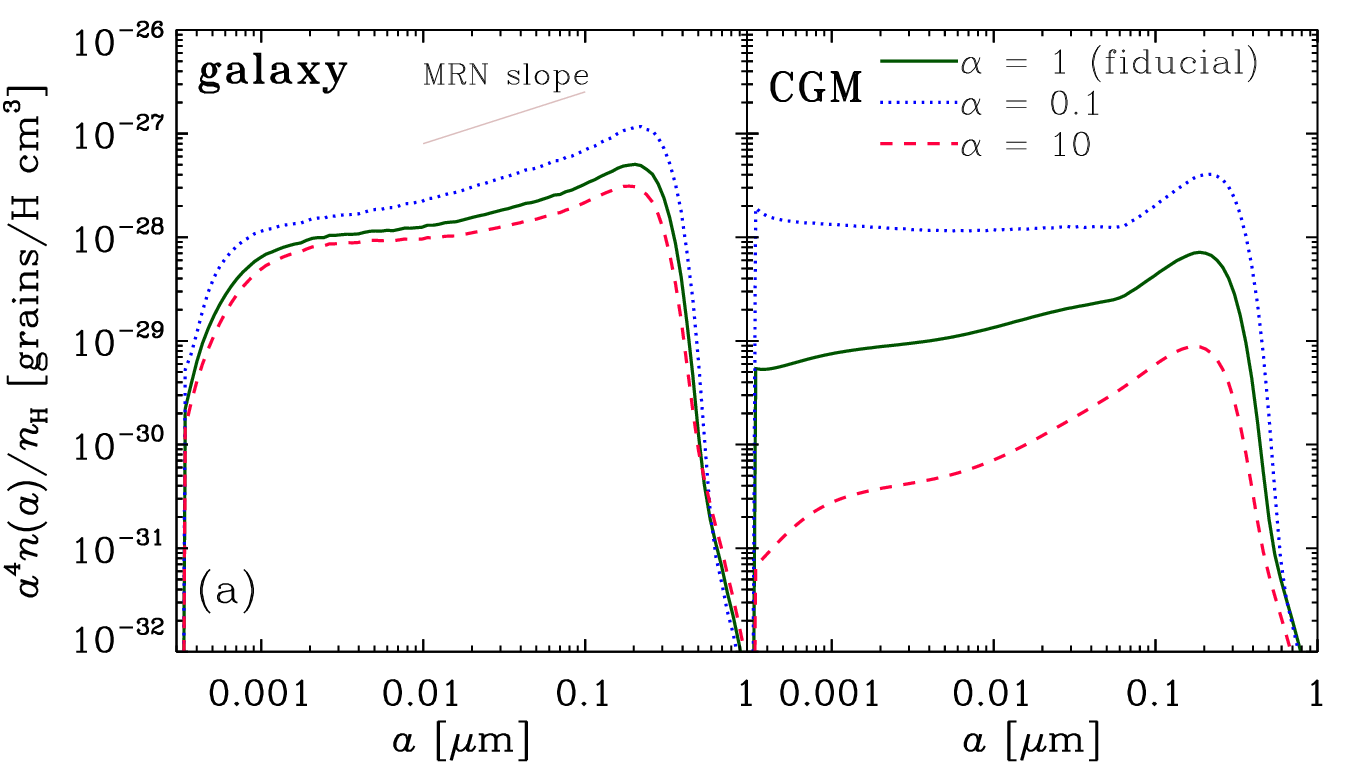}
 \includegraphics[width=\columnwidth]{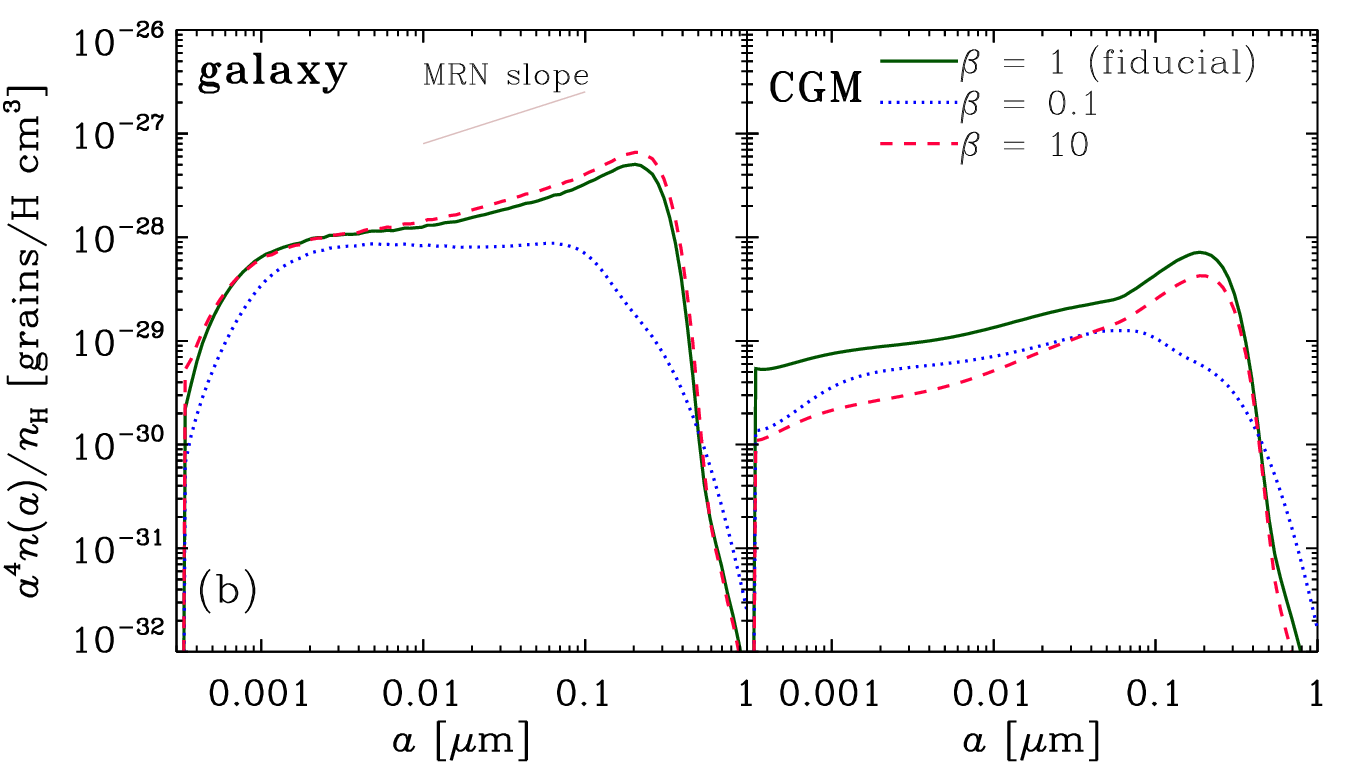}
 \caption{
 Grain size distributions at $t=10$~Gyr for various values for the inflow and outflow parameters:
 The solid, dotted, and dashed lines show the results for
 (a) $\alpha =1$ (fiducial), 0.1, and 10, respectively, and (b) $\beta =1$ (fiducial), 0.1, and 10,
 respectively. The grain size distributions and the MRN
slope are plotted in the same way as in figure \ref{fig:gsd}.}
 \label{fig:inflow}
\end{figure}

In figure \ref{fig:inflow}(a), we show the results for $\alpha =0.1$--10 at $t=10$~Gyr
with the other parameters fixed to the fiducial values (table \ref{tab:param}).
To simplify the presentation, we only compare the results at $t=10$ Gyr, but
the following differences caused by the changed parameter ($\alpha$ in the case here)
qualitatively apply to other ages.
In the galaxy, a lower inflow rate (larger $\alpha$) leads to
slower metal and dust enrichment.
Accordingly, the overall dust abundance in the CGM is also lower
for larger $\alpha$ (subsection \ref{subsec:mass}).
Different slopes in grain size distribution for different $\alpha$
are clear both in the galaxy and in the CGM.
In the galaxy, larger $\alpha$ leads to slower coagulation
because of lower dust abundance; thus,
the abundance of large grains is slightly suppressed
for larger $\alpha$. In contrast, the grain size distribution in the CGM shows more dominance
of large grains for larger $\alpha$. This is explained by different shattering efficiencies:
A low dust abundance in the CGM leads
to less efficient dust processing by shattering
because of a low grain--grain collision rate, leading to less production of small grains.
This explains the more large-grain-dominated
grain size distribution in the CGM for larger $\alpha (=10)$.

In figure \ref{fig:inflow}(b), we investigate the effects of outflow by showing the results
for $\beta =0.1$--10 at $t=10$ Gyr.
In the galaxy, the overall dust abundance, especially at large grain radii,
is lower for both small and large $\beta$. If $\beta$ is small,
the galaxy loses gas easily
so that the chemical enrichment does not proceed efficiently.
The resulting metallicity is about 1/3 solar for $\beta =0.1$, and as a result
the dust abundance is
low. Therefore, coagulation, which is the efficient formation path of large grains,
is less efficient for small $\beta$.
Accordingly, dust in the CGM is less abundant, especially at large grain radii.
In the case of large $\beta (=10)$,
the grain size distribution, including the overall dust abundance, in the galaxy is
similar to that in the fiducial case ($\beta=1$). This is because
the gas mass is maintained in the galaxy well if $\beta$ is larger than 1
(i.e., the effect of outflow is moderate).
The dust abundance is lower in the CGM in the case of
$\beta =10$ than in the fiducial case because the dust
supply from the galaxy is less efficient.

\subsection{Effects of dust processing in the CGM}\label{subsec:sput_shat}

In the CGM, two dust processing mechanisms are considered: sputtering and shattering.
As shown in table \ref{tab:param}, we change
$\tau_\mathrm{sput}$ and $n_\mathrm{H,cool}^\mathrm{C}$ to regulate sputtering and
shattering, respectively. We concentrate on the grain size distributions at $t=10$ Gyr
as we did in the previous subsection.

\begin{figure}
 \includegraphics[width=\columnwidth]{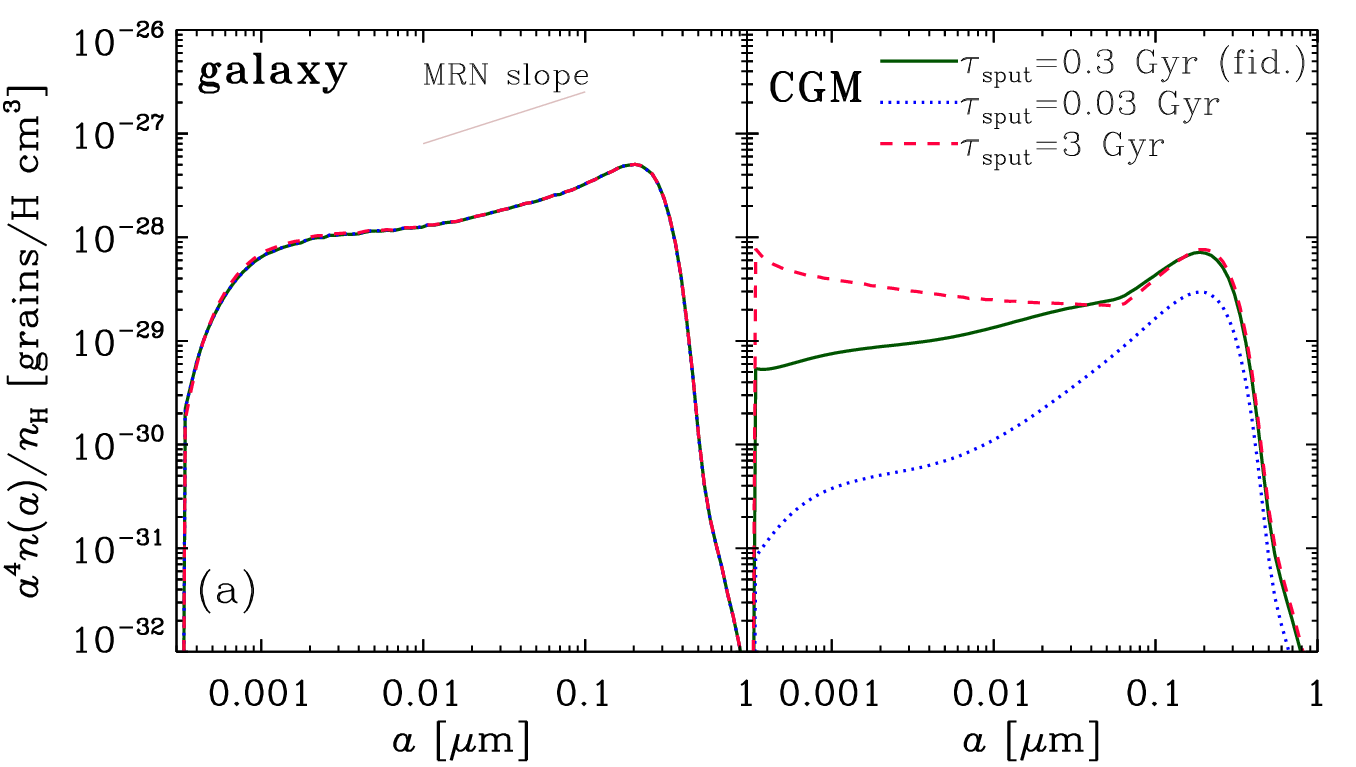}
 \includegraphics[width=\columnwidth]{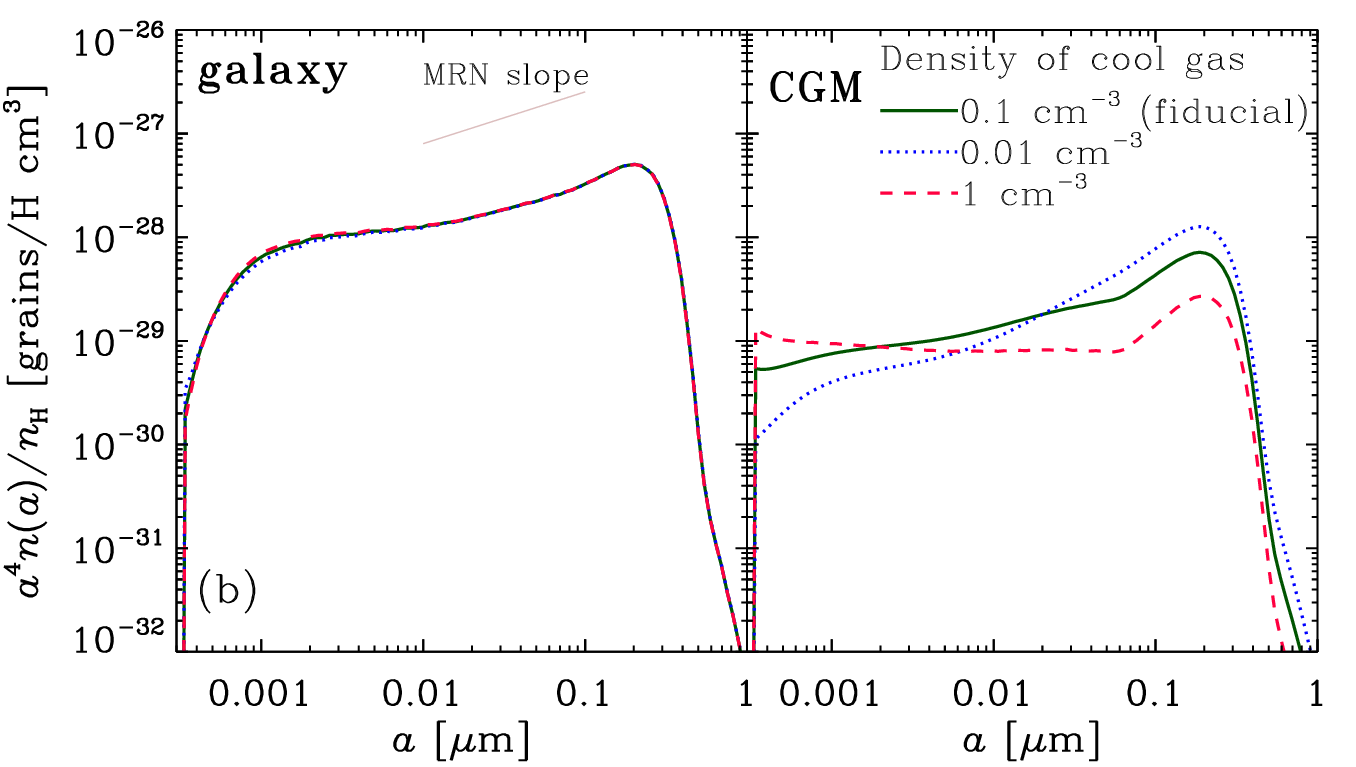}
 \caption{
 Same as figure \ref{fig:inflow} but for various sputtering and shattering
 parameter values:
 (a) $\tau_\mathrm{sput} =0.3$ (fiducial), 0.03, and 3 Gyr and
 (b) $n_\mathrm{H,cool}^\mathrm{C} =0.1$ (fiducial), 0.01, and 1 cm$^{-3}$
 for the solid, dotted, and dashed lines, respectively.
 }
 \label{fig:sput}
\end{figure}

In figure~\ref{fig:sput}(a), we show the grain size distributions for
$\tau_\mathrm{sput}=0.03$--3 Gyr. The grain size distribution in the galaxy is
little influenced by the sputtering parameter, while that
in the CGM is strongly affected.
As expected, the overall grain abundance is higher for longer $\tau_\mathrm{sput}$
because less dust is destroyed. The functional shape of the grain size distribution in the CGM
is also strongly affected by $\tau_\mathrm{sput}$. In particular, in the case of inefficient
sputtering ($\tau_\mathrm{sput}=3$ Gyr), the tail toward small grain radii is developed
because more small grains produced by shattering can survive.
Thus, the efficiency of sputtering affects
not only the overall grain abundance level but also the grain size distribution in the CGM.

In figure \ref{fig:sput}(b), we show the effects of shattering efficiency
by changing $n_\mathrm{H,cool}^\mathrm{C}$.
We observe that shattering in the CGM does not affect the grain size distribution in the galaxy.
In the CGM, a large value of
$n_\mathrm{H,cool}^\mathrm{C}$ means more efficient shattering,
which leads to more fragmentation of large grains. Thus, the abundance of large grains
is lower for larger $n_\mathrm{H,cool}^\mathrm{C}$.
Moreover, since smaller grains are more efficiently destroyed by sputtering,
more production of small grains by shattering leads to more efficient sputtering.
This interplay between shattering and sputtering decreases
the total dust abundance as observed in the case of $n_\mathrm{H,cool}^\mathrm{C}=1$
cm$^{-3}$ in figure \ref{fig:mass_sput}(b).

\section{Discussion}\label{sec:discussion}

\subsection{Reddening curves}\label{subsec:reddening}

For the purpose of an observational test,
we use the reddening curves observed for a sample of Mg \textsc{ii} absorbers,
which are considered to trace a gas component in the CGM
(see the Introduction).
We calculate the reddening curve using the method described
in section 2.4 of \citet{Hirashita:2021ab} (see also \cite{Hirashita:2020ab}).
{Because the observed reddening curves were derived from a large
galaxy sample, we assume that they reflect an averaged property of galaxies
to which our models with the above ranges of the parameters are applicable.}
We provide a brief summary for the
calculation procedure in what follows.

We define the reddening as the relative dust extinction
at two wavelengths. The extinction at wavelength
$\lambda$, $A_\lambda$ is described in units of magnitude, and is estimated as
\begin{eqnarray}
A_\lambda =1.086\kappa_\mathrm{ext}(\lambda )\mu m_\mathrm{H}N_\mathrm{H}\mathcal{D},
\end{eqnarray}
where $\kappa_\mathrm{ext}(\lambda )$ is
the mass extinction coefficient as a function of wavelength $\lambda$, and
$N_\mathrm{H}$ is the column density of hydrogen nuclei.
We calculate the reddening curve, that is, $A_\lambda -A_{\lambda_0}$
as a function of $\lambda$ with $\lambda_0$ being the reference wavelength.
The mass extinction coefficient is
estimated as
\begin{eqnarray}
\kappa_\mathrm{ext}=
\frac{\int_0^\infty\pi a^2Q_\mathrm{ext}(\lambda ,\, a)n(a)\,\mathrm{d}a}
{\int_0^\infty\frac{4}{3}\pi a^3sn(a)\,\mathrm{d}a},
\end{eqnarray}
where $Q_\mathrm{ext}(\lambda ,\, a)$ is the extinction efficiency, and is calculated using
the Mie theory \citep{Bohren:1983aa}.
The necessary grain property data are taken from
astronomical silicate (\cite{Weingartner:2001aa} and references therein)
and amorphous carbon (ACAR; \cite{Zubko:1996aa}).
We also adopt $s=3.5$ and 1.8 g cm$^{-3}$ from these papers
for astronomical silicate and amorphous carbon, respectively.
We do not use graphite, whose 2175 \AA\ feature is not clearly seen in
the actual observational data of Mg \textsc{ii} absorbers \citep{Hirashita:2021ab}.
As shown by \citet{Hirashita:2021ab}, while silicate explains more easily the rise of
extinction toward shorter wavelengths, carbonaceous dust fits better to the overall level of
extinction. Thus, following their approach, we mix silicate and amorphous carbon with a
mass ratio of 0.54:0.46, which is originally based on a Milky Way model
\citep{Hirashita:2019aa}.
However, the following results are not sensitive to the particular choice of the ratio
as long as the fractions of silicate and carbonaceous dust are comparable.

We adopt $\lambda_0=i/(1+z)$, where $i$ is the $i$-band wavelength (0.76~$\micron$).
We use the SDSS $u$, $g$, $r$, and $i$-band data for Mg\,\textsc{ii} absorbers at $z=1$ and 2 in
\citet{Menard:2012aa}. We also apply a range of
$N_\mathrm{H}=10^{19}$--$10^{20}$~cm$^{-2}$ for the typical column density of
an Mg\,\textsc{ii} absorber \citep{Lan:2017aa}.
The typical dust-to-gas ratio of Mg\,\textsc{ii} absorbers is 60--80 per cent of
the Milky Way value \citep{Menard:2012aa}; this high level is only achieved
at $t\sim 10$~Gyr in our model. Although the cosmic ages at $z=1$ and 2 are younger
than 10 Gyr, a smaller value of $\tau_\mathrm{SF}$ would give a similarly
high dust-to-gas ratio at a younger age. Because of this degeneracy between
$\tau_\mathrm{SF}$ and $t$, we use the results at $t=10$ Gyr to focus on
the most dust-enriched case achieved in our model.

\begin{figure}
 \includegraphics[width=0.95\columnwidth]{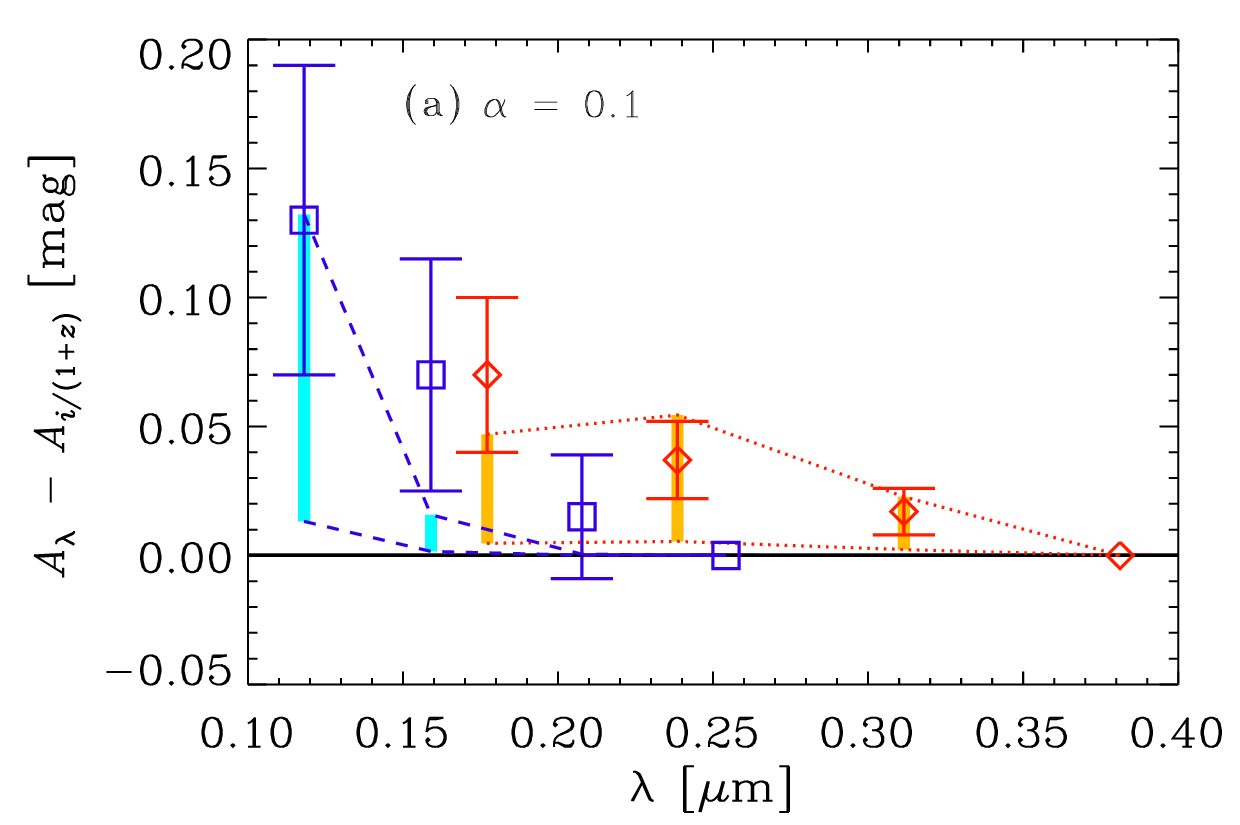}
 \includegraphics[width=0.95\columnwidth]{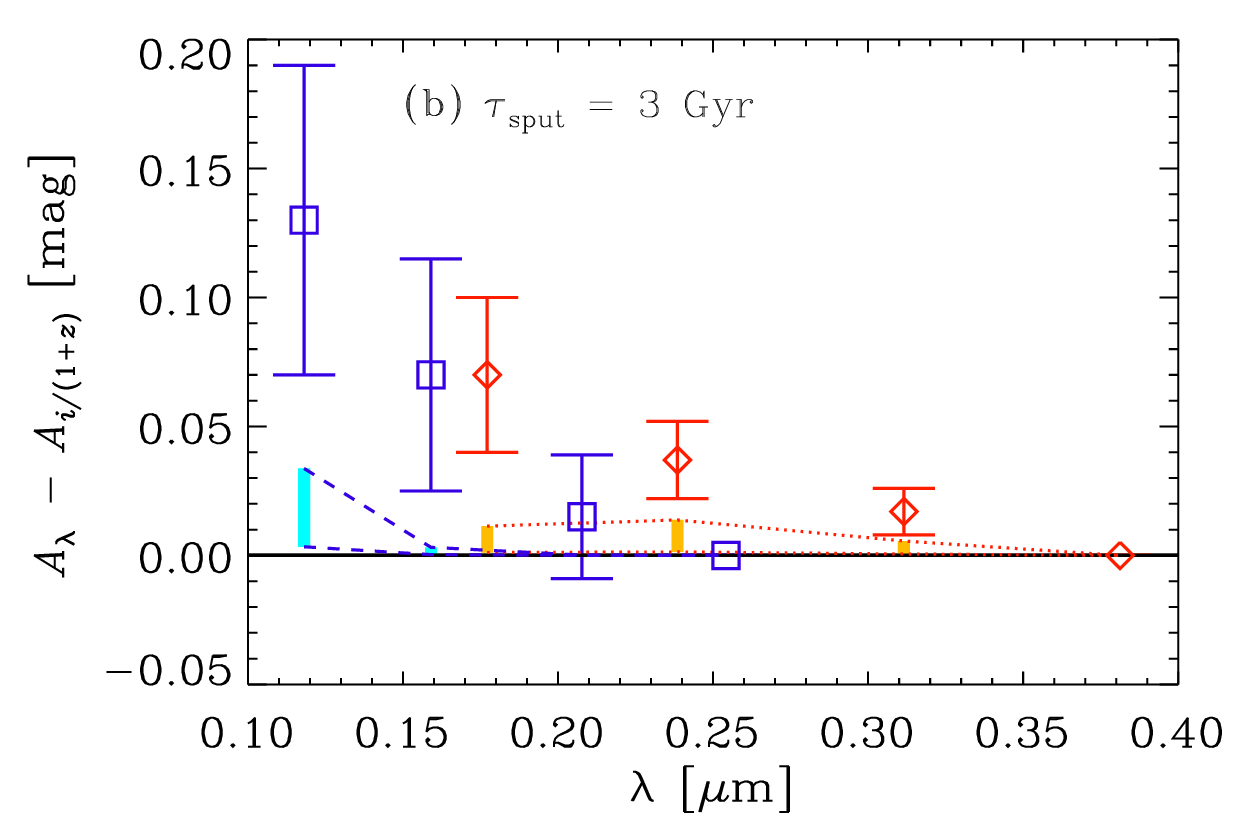}
 \caption{
Reddening curves (i.e.\ extinction relative to the $i$-band extinction)
at $z=1$ and 2
for the models with (a) $\alpha =0.1$, and (b) $\tau_\mathrm{sput}=3$~Gyr.
The other parameters are fixed to the fiducial values.
The wavelength is shown in the rest frame.
We adopt the grain size distribution at $t=10$ Gyr as a representative age when the central
galaxy is enriched up to solar metallicity in our model.
The orange and light blue bars, connected by the lines in the same color, present the
ranges corresponding to $N_\mathrm{H}=10^{19}$--$10^{20}$ cm$^{-2}$
in the SDSS $u$, $g$, and $r$ bands at $z=1$ and 2,
respectively, in our calculations. We also show the observational data of
Mg\,\textsc{ii} absorbers at $z=1$ and 2 (red diamonds and blue squares, respectively, with error bars)
taken from \citet{Menard:2012aa}, with the errors
expanded by a factor of 3 for a conservative comparison.
 }
 \label{fig:reddening}
\end{figure}

We find (not shown) that the fiducial model underpredicts the reddening by an order
of magnitude. Thus, we only show the cases where the grain abundance in the CGM
is enhanced; that is, small $\alpha$ (efficient inflow) and long $\tau_\mathrm{sput}$
(inefficient sputtering) as shown in subsection~\ref{subsec:mass}.
{In particular, sputtering cannot be stronger than the fiducial model.
Stronger sputtering also leads to more efficient destruction of small grains than large ones,
so that the resulting reddening curves would be too flat.}
In figure~\ref{fig:reddening}, we show the results for
$\alpha =0.1$ (Panel a) and $\tau_\mathrm{sput}=3$~Gyr (Panel b)
with the other parameters fixed to the fiducial values.
We observe that the model with efficient inflow
could explain the observed reddening while that with inefficient sputtering underpredicts it.
In fact, even the fiducial model is inefficient in destroying large
grains by sputtering (figure \ref{fig:sput}a),
so that the effect of weakening sputtering is
limited.
In contrast, an efficient inflow enriches the CGM to a high dust-to-gas ratio
not only because the galaxy is
more efficiently metal-enriched but also because the relatively small gas mass in the CGM
makes the dust enrichment easier (subsection \ref{subsec:mass}).
Therefore, a strong inflow whose time-scale is an order of magnitude shorter than the
star formation and outflow time-scales is favoured in explaining the level of
reddening observed for the CGM.

The above broad success in explaining the reddening curves in Mg \textsc{ii} absorbers
means that our evolution model of
grain size distribution is capable of explaining the actually observed optical properties of dust
in the CGM within reasonable parameter ranges.
We also emphasize that the steepness of the observed reddening curves toward
shorter wavelengths is naturally explained by these models that show enhanced
small grain abundances (especially, the model with $\alpha =0.1$ as shown in
figure \ref{fig:inflow}a).

\subsection{Possibilities of further improvement}

Our model treats each of the galaxy and the CGM as a one-zone object, neglecting
spatial variations in each zone. We further assume a fixed ratio for the dense and diffuse phases
and neglect possible time variations for the time-scales or efficiencies of
inflow, outflow, and dust processing.
Although this simplification serves to clarify the physical role of each process involved,
time variations caused by hydrodynamical evolution of the ISM and CGM may be important for
the efficiencies of various dust evolution processes.

Hydrodynamic simulations provide a useful platform where we treat dust evolution
in a manner consistent with the hydrodynamic evolution of galaxies
(e.g., \cite{Bekki:2013aa,McKinnon:2016aa,Aoyama:2017aa}).
In particular, \citet{Hou:2017aa} implemented a two-size grain model in
their isolated-galaxy simulations,
and showed that the grains in the CGM (out of the galactic disk) is biased to larger sizes.
Since they did not include sputtering in the CGM, the dominance of large grains in the CGM
should be due to an effect other than sputtering.
They interpreted this result as due to stellar feedback, which tends to eject
large grains produced by stars before they are affected by interstellar processing.
This effect cannot be included in our model, since we assume that the dust produced by stars
is instantaneously mixed with the preexisting dust in the galaxy.
\citet{Aoyama:2018aa} included sputtering in their cosmological hydrodynamic simulation,
and showed that the dust-to-metal ratio declines toward large galactocentric radii in the CGM.
They argued that their radial profile of
dust surface density is consistent with those derived from reddening observations
by \citet{Menard:2010aa} and \citet{Peek:2015aa}.
In fact, because of the uncertainty in the observational data, it is difficult to
constrain the sputtering process from the observations.
\citet{Zu:2011aa}, assuming a constant dust-to-metal ratio in a cosmological simulation,
also reproduced the radial profile of dust surface density.
The dust-to-metal ratio in the CGM they adopted is $\sim 0.2$,
which is larger than our fiducial case, but is well reproduced by the model with $\alpha\sim 0.1$,
and/or with $\tau_\mathrm{sput}\sim 3$~Gyr (figures~\ref{fig:mass_inflow}
and \ref{fig:mass_sput}). These models (especially the one with $\alpha\sim 0.1$) are
favored by the observed reddening curves
(subsection \ref{subsec:reddening}).

Shattering in the CGM is still challenging to include in hydrodynamic simulations because of
limited spatial resolutions. Since shattering occurs in cool clumps whose typical size is
$\sim 30$ pc \citep{Lan:2017aa}, spatial resolution is usually not enough in galaxy-scale
hydrodynamic simulations. The model in this paper could be used
for a subgrid model to be implemented in such simulations.

{Hydrodynamic simulations could also be used to determine the free parameters
(table \ref{tab:param}) adopted in this paper. In reality, they could have dependence on
the galaxy mass. Also, the galaxy mass assembly history is important to include.
Cosmological hydrodynamic simulations would provide a suitable platform on which we
investigate these issues.}

\section{Conclusions}\label{sec:conclusion}

We investigate the evolution of grain size distribution in the CGM by constructing a model
that describes the chemical enrichment in the galaxy and the CGM. We include
the mass exchange between the galaxy and the CGM by galactic inflows and outflows.
We also utilize the evolution model of grain size distribution developed in
our previous papers and extend it to describe the CGM dust. For dust evolution processes,
we include stellar dust production, SN destruction, shattering, accretion, and coagulation for
the galaxy, while we consider sputtering and shattering in the CGM.

We first present the evolution of the dust abundance in the CGM, and find broadly
consistent results with OH24 for the dependences on the processes
(inflow, outflow, and sputtering) that were also included in their dust mass
evolution model. We tend to underestimate the dust abundance in the CGM compared with OH24, which is
attributed to different assumptions on the mass evolution in the CGM.
We newly find that shattering in the CGM has an appreciable influence on the dust abundance
because small grains produced by shattering are easily destroyed by sputtering.

After these predictions, we show our main results, that is,
the evolution of grain size distribution.
We confirm that the evolution of grain size distribution in the galaxy is similar to our previous
models that treated the galaxy as a closed box.
Therefore, the inclusion of the CGM does not significantly affect the dust evolution in the galaxy.
The functional shape of the grain size distribution in the CGM to some extent follows that
in the galaxy, indicating that the grain properties in the CGM are strongly affected by interstellar
dust processing.
We also find, however, that the slope of the grain size distribution in the CGM
is sensitive to dust processing in the CGM, that is, sputtering and shattering.
If sputtering is dominant over shattering
as is realized in the cases for inefficient
dust enrichment with large $\alpha$, for efficient sputtering with short $\tau_\mathrm{sput}$, or
for inefficient shattering with small $n_\mathrm{H,cool}^\mathrm{C}$,
the grain size distribution in the CGM is more biased to large grains compared with that in
the galaxy.
In contrast, if shattering is dominant over sputtering as in the
opposite case for $\alpha$, $\tau_\mathrm{sput}$ or $n_\mathrm{H,cool}^\mathrm{C}$,
the grain size distribution
is more biased to small sizes compared with that in the galaxy.

To further predict observable features of the CGM dust, we
examine the reddening curve.
We find that our fiducial model underpredicts the reddening observed for a large
sample of background quasars by an order of magnitude.
This underprediction is effectively resolved by efficient inflow mainly because of higher
dust abundance achieved.
The steepness of the reddening curve is also consistent with the case of efficient inflow
since the small-grain abundance is enhanced.

Our results in this paper gives a basis on which the evolution of grain size distribution
in more sophisticated calculations
such as hydrodynamic simulations is interpreted, or
on which subgrid models in hydrodynamic simulations are developed.
The grain size distributions in the CGM calculated in this paper
are also essential in
predicting the dust extinction in the CGM, which could cause a systematic
bias for the colors of background objects.
{Our model is also general enough to be applied to various populations of galaxies
in the nearby Universe as well as at high redshift. Future studies focused on a particular
type of galaxies using our model would be useful if we choose an appropriate set of
the relevant parameters governing the dust enrichment in the CGM.}

\begin{ack}
{We are grateful to the referee, Shohei Aoyama, for his useful comments that improved the
discussions and presentations in this paper.}
We thank the National Science and Technology Council for support through grant
111-2112-M-001-038-MY3, and the Academia Sinica for Investigator Award AS-IA-109-M02.
\end{ack}




\bibliographystyle{aa}
\bibliography{/Users/hirashita/bibdata/hirashita}

\end{document}